\documentclass[letterpaper]{article} % DO NOT CHANGE THIS
\usepackage{aaai25}  % DO NOT CHANGE THIS
\usepackage{times}  % DO NOT CHANGE THIS
\usepackage{helvet}  % DO NOT CHANGE THIS
\usepackage{courier}  % DO NOT CHANGE THIS
\usepackage[hyphens]{url}  % DO NOT CHANGE THIS
\usepackage{graphicx} % DO NOT CHANGE THIS
\usepackage{natbib}  % DO NOT CHANGE THIS AND DO NOT ADD ANY OPTIONS TO IT
\usepackage{caption} % DO NOT CHANGE THIS AND DO NOT ADD ANY OPTIONS TO IT
\usepackage{algorithm}
\usepackage{algorithmic}
\usepackage{subfig}
\usepackage{xcolor}
\usepackage{newfloat}
\usepackage{listings}
\DeclareCaptionStyle{ruled}{labelfont=normalfont,labelsep=colon,strut=off} % DO NOT CHANGE THIS
\floatstyle{ruled}
\newfloat{listing}{tb}{lst}{}
\floatname{listing}{Listing}
\title{Auditing Yelp's Business Ranking and Review Recommendation \\ Through the Lens of Fairness}
\author{
    Mohit Singhal\textsuperscript{\rm 1}\thanks{Work done at The University of Texas at Arlington.}, Javier Pacheco\textsuperscript{\rm 2}, Seyyed Mohammad Sadegh Moosavi Khorzooghi\textsuperscript{\rm 2}, Tanushree Debi\textsuperscript{\rm 2}, Abolfazl Asudeh\textsuperscript{\rm 3}, Gautam Das\textsuperscript{\rm 2}, Shirin Nilizadeh\textsuperscript{\rm 2}
}
\affiliations{
    \textsuperscript{\rm 1}Northeastern University\\
    \textsuperscript{\rm 2}The University of Texas at Arlington\\
    \textsuperscript{\rm 3}University of Illinois Chicago\\

    m.singhal@northeastern.edu, (jxp6862,seyyedmohammads.moosavikhorzoog,txd8597)@mavs.uta.edu, asudeh@uic.edu, gdas@cse.uta.edu, shirin.nilizadeh@uta.edu

}

\begin{document}

\maketitle

\begin{abstract}
Auditing is critical to ensuring the fairness and reliability of decision-making systems. However, auditing a black-box system for bias can be challenging due to the lack of transparency in the model's internal workings. In many web applications, such as Yelp, it is challenging, if not impossible, to manipulate their inputs systematically to identify bias in the output. 
Yelp connects users and businesses, where users identify new businesses and simultaneously express their experiences through reviews. 
Yelp recommendation software moderates user-provided content by categorizing it into recommended and not-recommended sections. The recommended reviews, among other attributes, are used by Yelp's ranking algorithm to rank businesses in a neighborhood.
Due to Yelp's substantial popularity and its high impact on local businesses' success, understanding the bias of its algorithms is crucial. 

This data-driven study, for the first time, investigates the bias of Yelp's business ranking and review recommendation system. We examine three hypotheses to assess if Yelp's recommendation software shows bias against reviews of less established users with fewer friends and reviews and if Yelp's business ranking algorithm shows bias against restaurants located in specific neighborhoods, particularly in hotspot regions, with specific demographic compositions. Our findings show that reviews of less-established users are disproportionately categorized as not-recommended. 
We also find a positive association between restaurants' location in hotspot regions and their average exposure. Furthermore, we observed some cases of severe disparity bias in cities where the hotspots are in neighborhoods with less demographic diversity or higher affluence and education levels. 
\end{abstract}
\begin{center}
\large
\textcolor{red}{\textbf{This paper has been accepted at ICWSM 2025, please cite accordingly.}
}\end{center}
\section{Introduction}
The biases of machine learning-based decision-making systems are a major concern.
These biases have been identified in various domains, including predicting recidivism against African Americans~\cite{probublica, dressel2018accuracy,flores2016false}, and discriminatory practices against women in job screening and online advertisements~\cite{sweeney2013discrimination}. These instances underscore the urgent need for systematic auditing to uncover and address biases, as the consequences of biased decisions can perpetuate societal inequalities and undermine the ethical foundations of these systems. 

However, many of these systems are black-box, wherein the training data, algorithms, and model parameters are not publicly available. This lack of transparency makes the task of auditing for bias more challenging~\cite{caton2020fairness,harrison2020empirical}. 
Prior works proposed methods to audit black-box classifiers and recommendation systems~\cite{adler2018auditing,datta2016algorithmic,feldman2015certifying,tan2018distill}.
These works either make repeated calls to the black-box model API~\cite{kim2019multiaccuracy} or remove, permute, or obscure some of the features, or mimic the model~\cite{adler2018auditing,tan2018distill}
However, in certain scenarios, ethical and design restrictions may prevent auditors from probing the black-box system with carefully crafted input data. 
Consequently, the sole available resource for investigating bias becomes the output data already generated by the system.
This paper, for the first time, audits one such web application, Yelp, specifically auditing its business ranking and review recommendation system. 

On Yelp, users identify businesses in a local city or neighborhood, connect to other users, and share their experiences about the businesses with other Yelp users through their reviews and ratings.  
Yelp uses an automated content moderation tool called Yelp recommendation software to filter certain reviews based on four key considerations: conflicts of interest, solicited reviews, reliability, and usefulness~\cite{yelp-reco}. 
These filtered reviews are not removed from the site but are moved into a separate, less visible section called \emph{Not Recommended}, shown as a page link at the end of the businesses' page~\cite{yelp-reco-help1}.
Reviews are important because other Yelp users rely on them to choose a business. They also affect the business's overall star rating, which can subsequently affect its ranking in the Yelp search results. 

\textbf{Hypothesis 1 on Yelp Review Recommendation (Filtering) system:} 
Yelp has a mechanism to encourage users to provide ``high-quality'' reviews. For example, active Yelp users with many reviews, friends, etc., who also constantly provide helpful reviews receive an \emph{Elite} badge~\cite{yelp-elee1}. 
Prior works have shown that \emph{elite} user reviews are given higher weightage~\cite{kamerer2014understanding,nilizadeh2019think}.
However, using these users' account characteristics for filtering their reviews can lead to bias against \emph{less established} users who have a lower no. of reviews and friends. Filtering these users' reviews can be very frustrating for them or even considered unfair to them, as pointed out by Eslamin et al.~\cite{eslami2019user}.
Quotes from their qualitative study show this frustration and disappointment: ``A few friends and a few more reviews will take you out of the filter algorithm and allow your reviews to be posted.'' 

Additionally, prior research has extensively explored the issue of \textit{long-tailed} item recommendation, where less prolific users or items with limited activity receive less attention~\cite{fleder2009blockbuster,silva2023user}.
Furthermore, studies have highlighted the adverse effects of unfair content moderation on user disengagement~\cite{duffy2022platform,singhal2023sok}. 
Therefore, the first hypothesis examines the bias of Yelp's recommendation system against a new type of sensitive group, i.e., \emph{less-established users} in contrast to \emph{well-established users}:
\emph{\textbf{H1.} Reviews written by users with fewer friends and reviews are more likely to be categorized as not recommended.}

\textbf{Hypotheses 2 and 3 on the Yelp's Business Recommendation System:} Yelp uses a black-box ranked-retrieval model~\cite{yelp-serar} to list businesses per users' search query, looking for a specific type of business, location, services, etc. 
The search result shows a ranked list of businesses on several pages. 
Businesses listed among the firsts on the first few pages are considered to have higher visibility and exposure, as some studies have shown users tend only to check the first or a few first pages~\cite{joachims2017accurately}. 
Studies have also shown that this ranking highly affects businesses because the higher visibility and exposure~\cite{singh2018fairness} means higher foot traffic and increased revenue~\cite{luca2016reviews}. Therefore, a fair ranking is essential for a business. 
Many factors, including location, can impact a business's ranking on Yelp search results. For example, some neighborhoods are hotspots for restaurants, and several businesses thrive in these locations. This can be due to being in the most touristy and visited parts of the city, which garner more foot traffic. 
This can, in turn, lead to a higher number of (positive) reviews for the businesses, {\em giving even more exposure to them.} The higher exposure, in turn, drags even more customers to these businesses, giving them a further boost in fame and popularity, even though they did not provide a better service or higher quality food than other businesses. 
This cycle of discrimination continuously increases the popularity and revenue gap between these businesses located in different neighborhoods. As a result, it causes  {\em ``the rich to get richer and the poor to get poorer''} as it is shown in prior works that there is a correlation between reviewer positiveness and restaurants being in a hotspot or popular location~\cite{kokkodis2020your,huang2016effects}. 
Hence, we hypothesize that:

\emph{\textbf{H2.} In Yelp's ranking of businesses in a city, businesses located in hotspots tend to have higher exposure.}

Furthermore, while the location itself might not be considered sensitive, it can be associated with sensitive attributes, such as neighborhoods' economic status and demographic and racial composition. Therefore, if Yelp's business ranking algorithm is biased against businesses that are not located in hotspots, and if these neighborhoods are associated with a specific demographic composition, then Yelp is (implicitly) biased against this specific population.   
Hence, we hypothesize that:

 \emph{\textbf{H3.} Hotspots are positively associated with less demographic diversity, high education levels, high income, and low unemployment neighborhoods}

To examine bias, we relied on popular definitions of fairness and bias, such as 
\emph{exposure} and used statistical tests, such as quantile linear and logistic regression, to identify associations and relationships between Yelp's outcomes and the sensitive groups. Since the inherent black-box structure and the lack of access to training data and algorithms make it challenging to audit Yelp, we designed two data collection frameworks to minimize bias. Note that in contrast to other auditing frameworks, due to ethical considerations, we do not generate inputs (create accounts or write reviews) to pass them through the system but instead analyze the existing data on the platform. 

To audit the algorithmic bias of the Yelp Review Recommendation (Filtering) system, we obtained \emph{recommended} reviews of 15K random businesses from Yelp Dataset Challenge in 2023~\cite{yelp-datas}. We also created a framework to obtain all the \emph{not-recommended} reviews of these businesses from Yelp as Yelp does not provide it in the Dataset Challenge. In total, we obtained 707K recommended and 178K not-recommended reviews. To audit the algorithmic bias of the Yelp Ranking system, we first obtained the ranking of all restaurants for 9 cities in the US using our custom-build framework as Yelp Dataset Challenge does not provide the ranking of businesses.
Through the statistical analysis, we found that reviews written by users with a higher number of friends and reviews are more likely to be categorized as \emph{recommended}, hence showing a disparate impact for new or less-active users ($p<0.001$). 
Our quantile analysis showed a statistically significant positive correlation between a restaurant's presence in a hotspot and its average exposure ($p<0.001$). This finding was observed across all nine cities, indicating that the Yelp algorithm disproportionately accords higher exposure to restaurants in the city's hotspots. Furthermore, we noted more severe cases of disparity bias in cities where these hotspots are placed in neighborhoods with less demographic diversity or areas with higher affluence and education levels. For example, in Chicago, we found that hotspot regions are usually in highly educated and wealthy neighborhoods. 
Hence, we can conclude that there would be bias against the restaurants in neighborhoods with lower wealth and less educated people. The association between hotspots and demographic composition can also be due to existing deep underlying societal biases. 

Our results highlight the importance of {\em responsible data science practices} in data-driven systems such as Yelp. Irresponsible ranking and recommendation models and algorithms cannot only create unfair results but can also create discriminatory feedback loops that keep segregating the cities and marginalizing people. 

\section{Related Work}

\textbf{Fairness \& Bias in Recommendation Systems.} 
Prior works have found that the recommendation systems incur biases~\cite{zhang2021fairness,ali2019discrimination,abdollahpouri2019unfairness,sanchez2020does}, some against a gender or race group~\cite{li2021towards,hannak2017bias,buolamwini2018gender,zhao2017men} leading to under-representation of some social groups and denial of economic opportunities for others~\cite{hannak2017bias,abdollahpouri2019multi,beutel2019fairness,koenecke2023popular}. 
For example, Ali et al.~\cite{ali2019discrimination} found that skewed delivery of ads occurred on Facebook leading to bias. Ekstrand et al.~\cite{ekstrand2017demographics} found that recommender systems suffer
of what is known as ``sample size bias.'' Edizel et al.~\cite{edizel2020fairecsys} found that bias that exists in the real world can be modeled and amplified by recommender systems. Marlin et al.~\cite{marlin2012collaborative} found selection bias in rating data collected via user survey. 
Recommender systems can also affect users decision making
process, known as decision bias, and Chen et al.~\cite{chen2013human} show how understanding this bias can improve recommender systems. Dozens of methods to make these ML systems fair are proposed~\cite{li2021towards,li2021user,fu2020fairness}. 

\textbf{Fairness \& Bias in Ranking.} 
Ranking fairness implies that comparable items or groups of items receive similar visibility. Prior works have extensively studied fairness in ranking~\cite{stoyanovich2018online,geyik2019fairness,patro2022fair,zehlike2021fairness,yang2019balanced}. Scholarships have examined probability-based fairness notions and provided mechanisms to design fair ranking~\cite{asudeh2019designing,geyik2019fairness,zehlike2017fa}.
For example, Asudeh et al.~\cite{asudeh2019designing} developed a system that assists individuals in choosing criterion weights that lead to significant fairness. Singh et al.~\cite{singh2018fairness} proposed a framework for formulating fairness constraints on rankings. 
In fair ranking tasks, a frequently studied problem is how to distribute the exposure opportunity to candidates fairly
~\cite{diaz2020evaluating,morik2020controlling,singh2018fairness,zehlike2020reducing}. Position bias is very common in recommendation systems~\cite{chen2023bias}. Popularity bias happens when
users to notice or interact with items in certain positions of lists with higher probability~\cite{collins2018study}. Maeve et al.~\cite{o2006modeling} shows that users often trust the first few results in the lists. Popularity bias is another form of ``exposure bias''~\cite{zheng2021disentangling}. Our work builds on these works by investigating the bias of Yelp's ranking system through the lens of exposure bias. 
\textbf{Yelp Recommendation System:} Mukherjee et al.~\cite{mukherjee2013yelp} investigated the factors that the Yelp review filtering algorithm uses, reporting that Yelp might filter reviews based on various behavioral features than linguistic ones. In our work, we examine the sensitive user-level features, such as  no. of friends and reviews. 
Kamerer~\cite{kamerer2014understanding} established that Yelp users who are prolific reviewers and are known and trusted by the community have their reviews in the recommended section. Yao et al.~\cite{yao2018yelp} found that reviews were most likely to be in the recommended section when conveying an overall positive message. Some works studied the opinions and reactions of users whose reviews were being removed~\cite{duffy2023platform,jhaver2019did}.For example, Eslami et al.~\cite{eslami2019user} found that users echoed that the system suppresses their voices.  
One relevant work to our study~\cite{amos2022reviews} empirically studied Yelp's review recommendation and found that reviews of businesses in lower-density and low-middle-income areas are most likely to be labeled as not recommended. Our study, however, investigates bias toward established and new users. 

\textbf{Bias in Content Moderation. } 
Scholarships have investigated bias and fairness in content moderation and how it affects the users~\cite{duffy2023platform,jhaver2019did,eslami2019user,singhal2023sok}, mostly focusing on its harm on marginalized communities via user studies~\cite{haimson2021disproportionate,seering2019moderator}. 
Eslami et al.~\cite{eslami2019user} found that elite users on Yelp defend the Yelp review recommendation algorithm because their reviews are rarely filtered. 

\section{Auditing Bias of Yelp’s Review
Recommendation (Filtering) System}
Researchers have proposed dozens of definitions of fairness for ML models~\cite{narayanan2018tutorial,ekstrand2022fairness,lazovich2022measuring,patro2022fair}. However, not all definitions of fairness can satisfy a specific use case. In this particular use case, we have three key stakeholders, i.e., Yelp, Yelp users, and Businesses on Yelp each having their own fairness perspective:
    \textbf{Yelp:} Yelp would want to ensure their review recommendation software has near-perfect accuracy, correctly identifying and highlighting honest and useful reviews. 
    Therefore, fairness to Yelp is to use any features or ML models that can provide such high performance, enhancing the experiences of both people and businesses who use the platform, and maintaining Yelp's popularity in the highly competitive market over other recommendation applications.  
    \textbf{Yelp Review Writers:} Review writers spend time and effort to create and maintain an account on Yelp and participate in the platform by connecting to other Yelp users and sharing their personal experiences with everyone. Therefore, fairness for these users can be acknowledging their opinions, making them available on the platform, and using them to rate the businesses. In contrast, if their reviews are placed in the not recommended section, they echo frustrations and call the system biased~\cite{eslami2019user}. 
 \textbf{Businesses on Yelp} would want Yelp to remove reviews that are low quality and have a high probability of being fake. Low quality and especially negative fake reviews can cause the star rating of a business to decrease, resulting in less foot traffic~\cite{hussain2019spam,horn2015business}.
   
These three perspectives can contrast each other. For example, to achieve high accuracy, the system might prefer to use the account's characteristics, such as the number of reviews and friends. This might result in the disproportional removal of reviews by less established users.

\subsection{Methodology and Bias Metrics} 
Prior works~\cite{koenecke2023popular,becerril2023method} have used statistical association between the systems' outcomes and the sensitive attributes to indicate bias. For example, Becerril~\cite{becerril2023method} used regression analysis to explain product prices and price differences as functions of consumer demographics.
To test \textbf{H1}, we used multivariate logistic regression to statistically investigate the relationship between the users' number of friends and reviews and their reviews being categorized as not recommended. We also controlled for various other confounding metrics for review quality and likelihood of being fake that could contribute to whether the review would be in the \emph{not recommended} section. Specifically, we controlled for the \emph{sentiment} of the review, which is a binary variable for either positive or negative review. Yao et al.~\cite{yao2018yelp} found that when the review is positive, it is more likely to be in the recommended section.
Additionally, we controlled for \emph{review length} and \emph{number of spelling mistakes} as these properties can be indicators of informative vs. low-quality reviews~\cite{kamerer2014understanding,mukherjee2013yelp}. 

\begin{figure} [h!]
    \centering
        \subfloat[Recommended]{ \includegraphics[width=0.5\columnwidth]{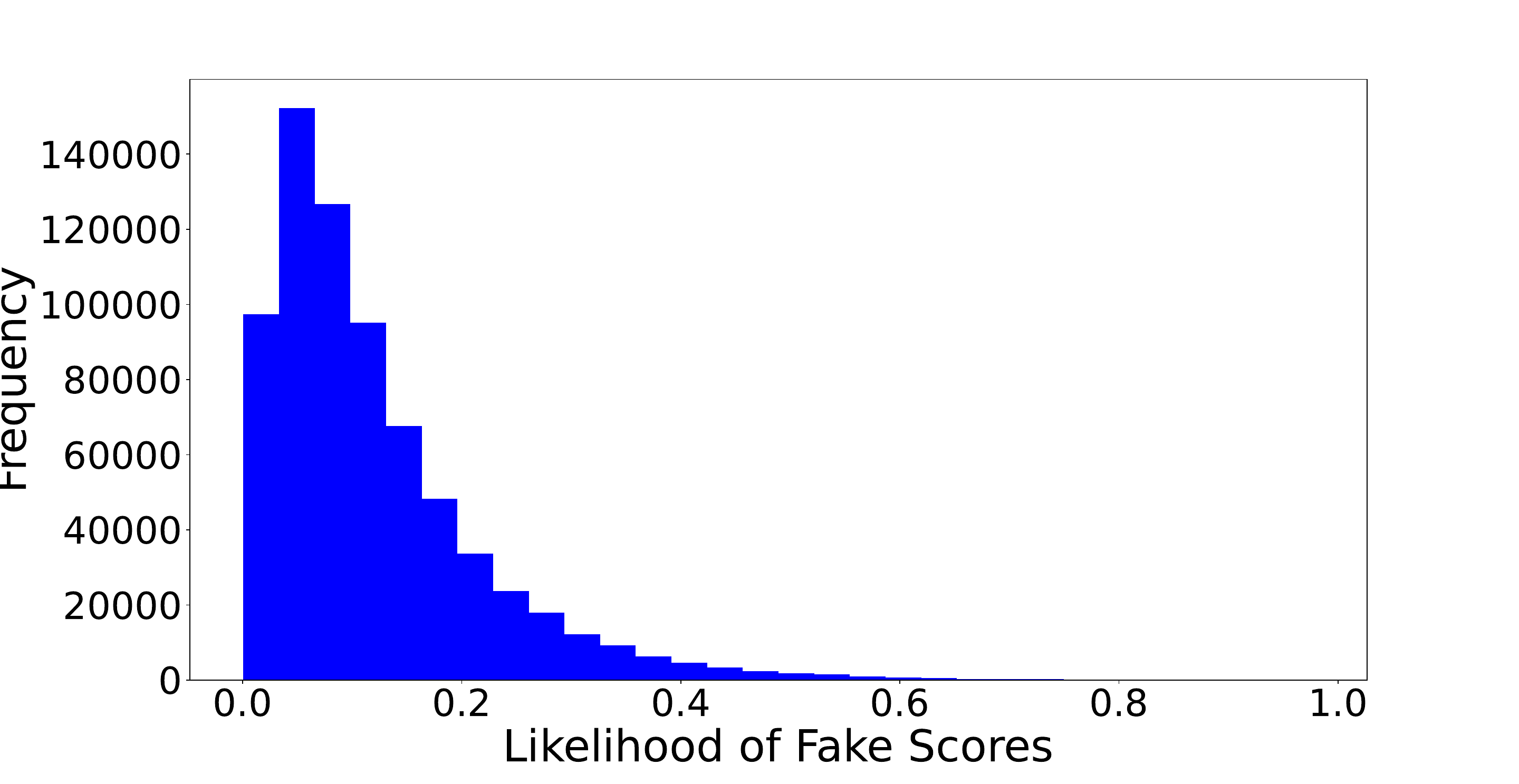}\label{fig:fig122}}
        \subfloat[Not Recommended]{\includegraphics[width=0.5\columnwidth]{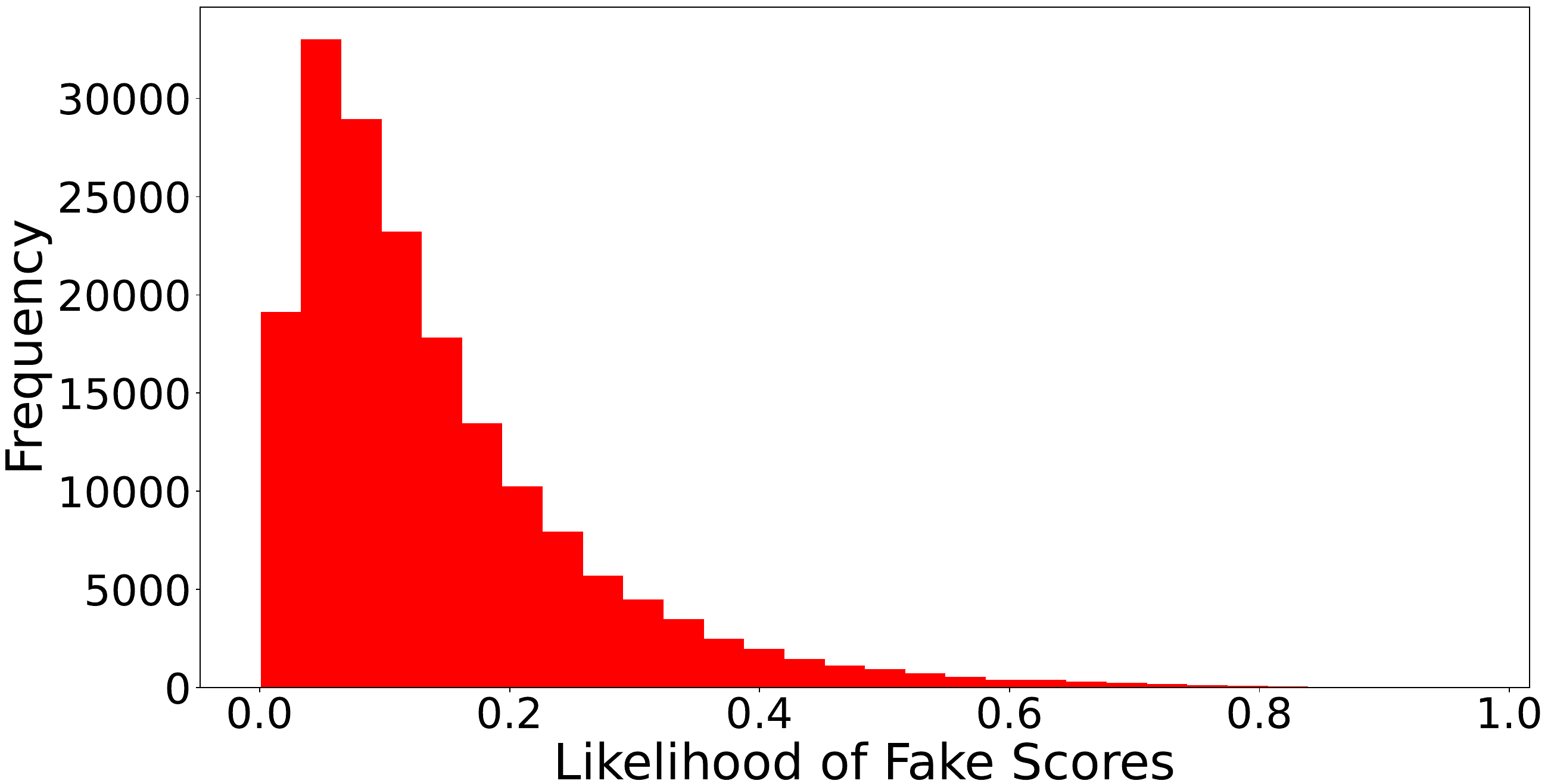}\label{fig:fig442}}
         \caption{Histogram of likelihood of reviews to be fake}
  \label{fig:fakee}
\end{figure}

\textbf{Fake review consideration.} Yelp has two subsections on the Not Recommended page, one for the not recommended reviews and one for those violating its community guidelines~\cite{yelptos}. We did not crawl the second section, hence our dataset does not have any reviews that are in the violation of Yelp ToS, including fake reviews. Hence, the biases in Yelp's fake review detection method minimally affect our analysis. The only concern is the false negatives that the method might have and the fake reviews that remain undetected in both sections on Yelp, i.e., recommended and not-recommended sections. We checked recent works such as~\cite{mohawesh2024fake,duma2024dhmfrd,ashraf2024leveraging}, that are about fake review detection and unfortunately, we could not identify any that open sources their code or data. Also, most of these methods include features extracted from (meta-) data or auxiliary data that we do not have access, making it almost impossible to implement them. To control for such cases, we developed a detection method. We first preprocessed the reviews by removing special characters, emoticons, etc., to minimize the sparsity of data~\cite{singhal2023cybersecurity}.
We extracted bi-grams and uni-grams from all the posts and considered the top 2000 with the highest TF-IDF values. Using the groundtruth dataset provided by~\cite{salminen2022creating}, we implemented a traditional SVM ML model and found an accuracy of 0.93 and an F1-score of 0.93, which is higher than reported in prior work~\cite{mukherjee2013fake,harris2019comparing}. Additionally, Figure~\ref{fig:fakee} shows the distribution of the likelihood of reviews being fake and interestingly, we can see that there is a high left skew, i.e., almost all reviews are benign. Hence, compared to this simple model, it seems Yelp is performing well in finding fake reviews. Hence we used this as a control variable in our model. 

Since the distribution of these variables is not normal but rather skewed, running the test on all the data cannot provide a full picture. We used quartile regression analysis~\cite{yu2003quantile}, which organizes data into three points—a lower quartile, median, and upper quartile—to form four groups of the dataset, each representing 25\% of the observations. 
Quartiles are examples of quantiles used in research as statistical quantities~\cite{yu2003quantile,koenker2004quantile}. 
After dividing the data into different quartiles, we run our regression model on each quartile separately. This allows us to interpret the results of each quartile separately. 
Regression models are valuable for analyzing how changes in one variable are associated with changes in another variable. In the context of bias, we used regression to understand how certain variables might be correlated with biased outcomes. However, it's essential to remember that correlation does not imply causation. A regression model can demonstrate a statistical relationship, but it cannot prove that one variable causes the bias. 

\subsection{Data collection} 
We developed a framework for collecting the datasets. 
First, we used the dataset provided by Yelp as part of their Yelp Dataset Challenge in 2023~\cite{yelp-datas} to identify and collect 707,658 \emph{recommended} reviews of about 144K businesses spanning over about 11 metropolitan cities in the USA. 
The Yelp Dataset Challenge does not provide the \emph{not recommended} reviews, and therefore, we deployed a custom-built crawler on Yelp to collect them. 
However, collecting such large-scale data is time-consuming and could generate much traffic, violating Yelp's Terms of Service.   
Therefore, we sampled 15K (10\%) of the businesses and obtained their not-recommended reviews. To avoid overwhelming Yelp with our requests, we delayed the data collection process in our crawler by 5 to 10 seconds, which was enough, and Yelp never blocked our requests.
We were able to obtain 178,747 not-recommended reviews for 13,239 businesses, as 1,761 of them did not have any non-recommended reviews at the time of data collection. 
Scraping the not recommended reviews, We obtained the full review text, review date, friend and review count, username, location, and the URL of the user profile image. However, it doesn't provide user IDs for them. Our code and data can be found and tested in this anonymous Dropbox link \url{https://tinyurl.com/pnt8uudb}. Our dataset does not include paid reviewers.
\begin{figure} [t]
    \centering
        \subfloat[Review Count]{ \includegraphics[width=0.45\columnwidth]{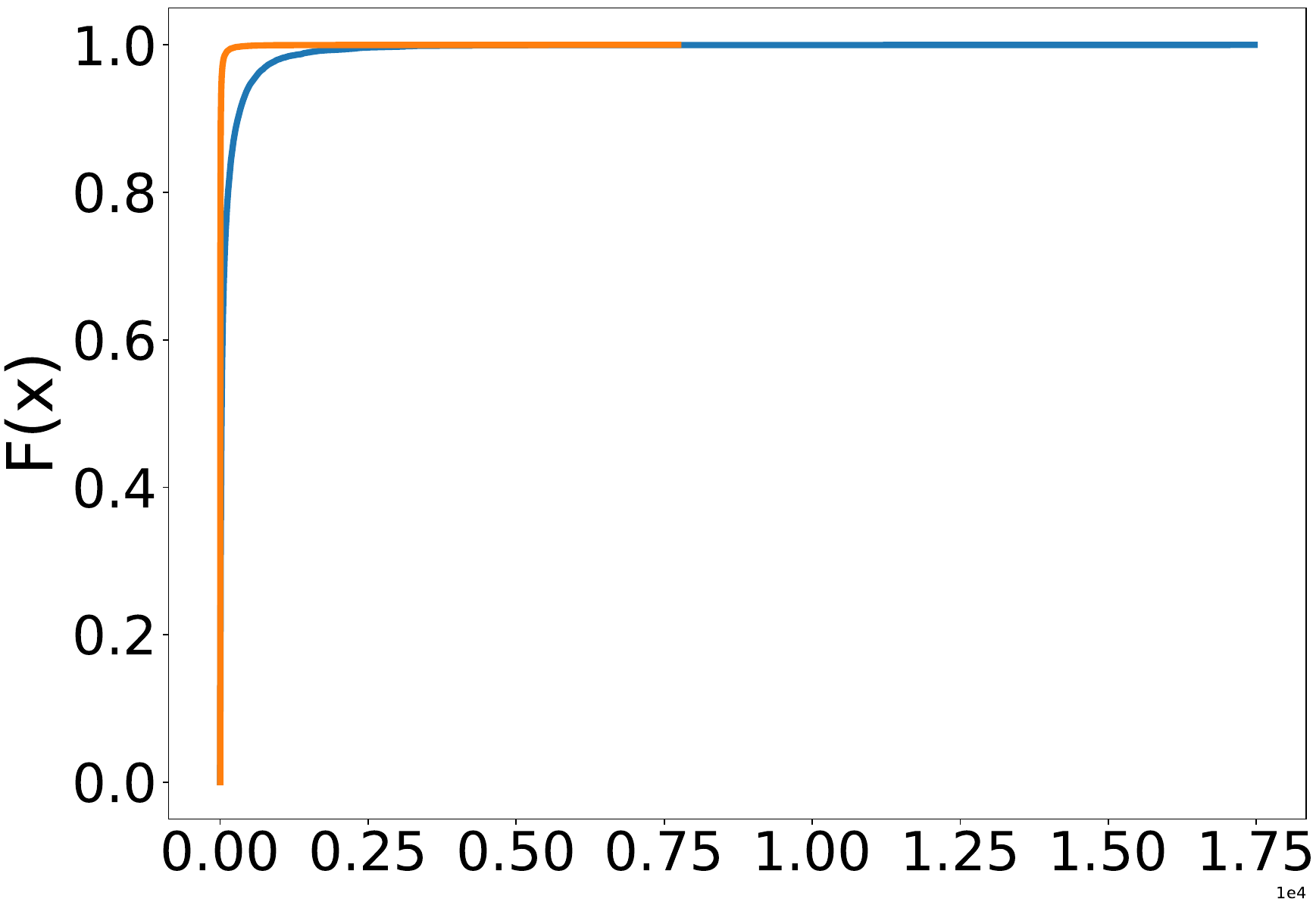}\label{fig:fig1}}
        % \hfill
        \subfloat[Friend Count]{\includegraphics[width=0.45\columnwidth]{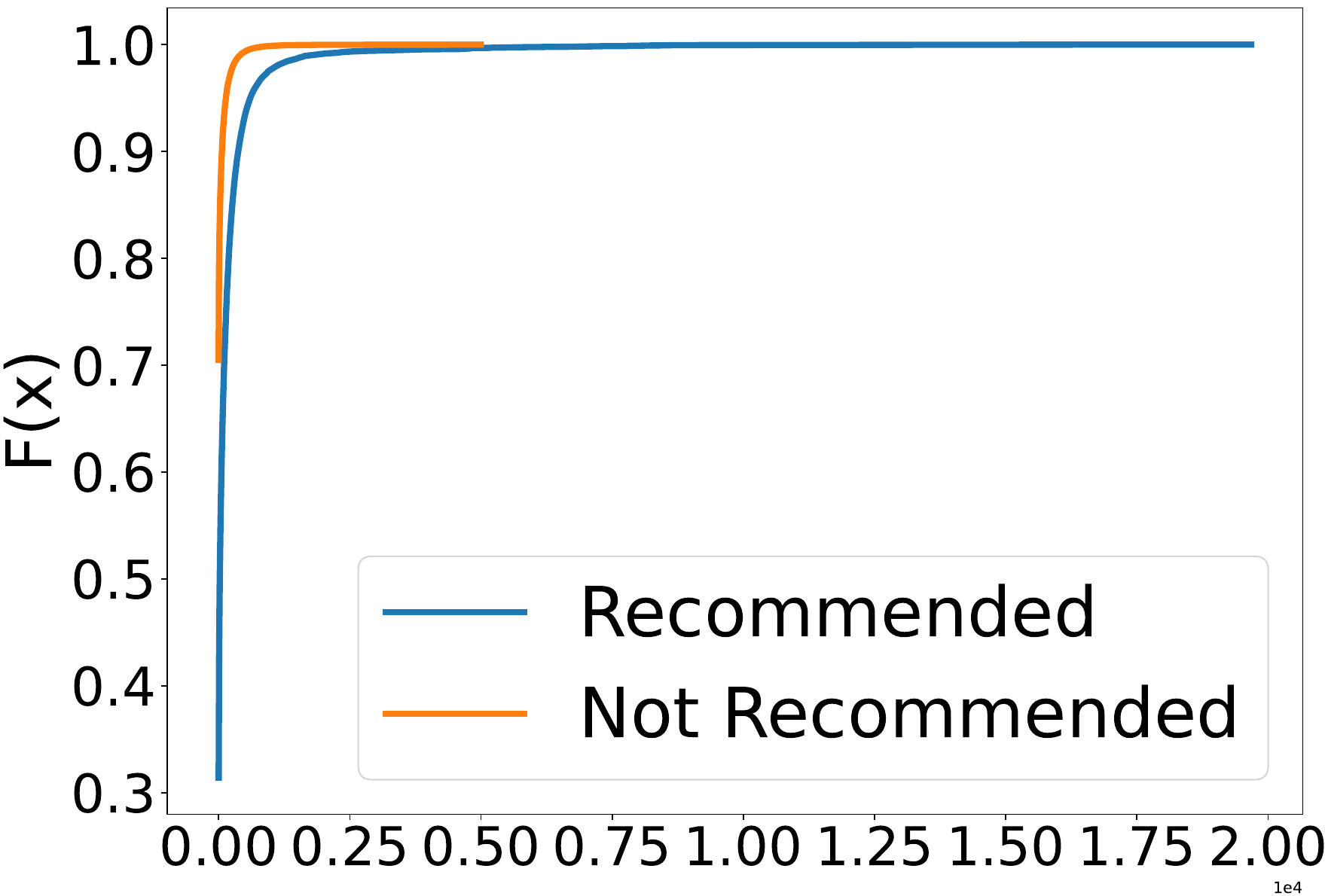}\label{fig:fig2}}
         \caption{CDF of Reviews and Friends for Recommended and Not Recommended Users}
  \label{fig:accounts}
\end{figure}
\begin{table}[ht!]
            \centering
            \caption{Descriptive statistics of dataset}
             \label{stat}
            \resizebox{\columnwidth}{!}{%
            \begin{tabular}{l|cccc|cccc}
            \hline  \\[-1.8ex] 
             &
             \multicolumn{4}{c}{Recommended Users } &
             \multicolumn{4}{c}{Not-Recommended Users} \\  
                 &     \multicolumn{4}{c}{(n=707,658)} &
             \multicolumn{4}{c}{(n=178,747)} \\   
             \hline         
             Variable &Min &Max &Mean &Med. &Min &Max &Mean &Med.  \\
             \hline          
           \# Friends &1 &19,679 &157.89 &21&0 &4,995 &23.79 &0 \\         
            \# Reviews &0 &17,473 &121.76 &24 &0 &7,738 &7.73 &2 \\          
            \hline
            \end{tabular}
            }
\end{table}

\textbf{Dataset Characterization: }
Table~\ref{stat} shows the descriptive statistics of the dataset. We could not find the unique users in the not recommended dataset, as Yelp does not provide user IDs of these users in that section. Therefore, we did not obtain the unique users in the recommended set as well. 
Figures~\ref{fig:fig1} and~\ref{fig:fig2} show the cumulative density function (CDF) of the number of friends and reviews, respectively. 
They show that these variables do not follow the normal distribution and on average, users in the recommended dataset have more friends and reviews than those in the non-recommended dataset. 

\begin{table*}[!tbh]
            \centering
            \caption{Results of the quartile regression}
            \resizebox{0.75\textwidth}{!}{%
            % \begin{tabular}{*{12}{|c}{|c|}}
            \begin{tabular}{lccccc}
            \hline  \\[-1.8ex] 
             & \multicolumn{5}{c}{Dependent variable: \textit{recommended} } \\  
             \hline         
             & Logistic & 0.25 Qnt.&0.50 Qnt. &0.75 Qnt. &1.00 Qnt.\\
             \hline          
            No. of Friends& 0.003(0.000)$^{***}$ &592.9 (0.997) &-0.453(0.000)$^{***}$&0.001(0.000)$^{***}$& 0.000(0.000)$^{***}$ \\ 
            No. of Reviews&0.049(0.000)$^{***}$&0.024(0.000)$^{***}$&0.024(0.000)$^{***}$&0.028(0.000)$^{***}$&0.011(0.000)$^{***}$\\
        Sentiment&-0.026(0.000)$^{***}$&-0.528(0.000)$^{***}$&-0.419(0.000)$^{***}$& -0.252(0.000)$^{***}$&-0.300(0.000)$^{***}$ 
    \\Spelling Mistakes&0.003(0.000)$^{***}$&0.031(0.000)$^{***}$&0.050(0.000)$^{***}$& 0.062(0.000)$^{***}$&0.062(0.000)$^{***}$ \\
            Length of Review&-0.000(0.000)$^{***}$&0.000(0.618)&-0.000(0.000)$^{***}$&-0.000(0.000)$^{***}$&-0.000(0.834) \\
            Likelihood of Fake&-1.576(0.000)$^{***}$&-1.651(0.000)$^{***}$&-1.893(0.000)$^{***}$&-1.964(0.000)$^{***}$&-2.073(0.000)$^{***}$ \\
            No. of Friends (IRR)& 1.003&3.182e+257 &0.635 &1.001 & 1.000 \\
            No. of Reviews (IRR)&1.051&1.024& 1.025&1.028&1.0117\\
            Sentiment (IRR)& 0.769&0.589 &0.657 &0.776 & 0.740 \\
            Spelling Mistakes (IRR)& 1.031&1.033 &1.051 &1.064 & 1.064 \\
            Length of Review (IRR)& 0.999&1.000 &1.000 &0.999 & 0.999 \\
            Likelihood of Fake (IRR)& 0.206&0.191&0.150&0.140&0.125\\
            \hline
            Observations& 886,405 &221,602&221,601&221,601&221,601\\
            \hline 
            No. of Reviews& &0.651(0.000)$^{***}$&0.152(0.000)$^{***}$&0.016 (0.000)$^{***}$&0.000(0.000)$^{***}$  \\ 
            No. of Friends&&0.004(0.000)$^{***}$&0.002(0.000)$^{***}$&0.000(0.000)$^{***}$&0.000(0.000)$^{***}$\\
            Sentiment&&-0.556(0.000)$^{***}$&-0.372(0.000)$^{***}$& -0.230(0.000)$^{***}$&0.165(0.009)$^{**}$ \\Spelling Mistakes&&0.002(0.024)$^{*}$&0.019(0.000)$^{***}$& 0.086(0.000)$^{***}$&0.080(0.000)$^{***}$ \\
            Length of Review&&0.000(0.000)$^{***}$&0.000(0.000)$^{***}$&0.001(0.000)$^{***}$&0.005(0.000)$^{***}$ \\
            Likelihood of Fake&&-1.557(0.000)$^{***}$&-1.950(0.000)$^{***}$&-2.233(0.000)$^{***}$&-2.011(0.000)$^{***}$\\
            No. of Reviews (IRR)& &1.917&1.164  & 1.016  &1.000  \\ 
            No. of Friends (IRR)&& 1.004&1.002&1.000&1.000\\
            Sentiment (IRR)& &0.573 &0.689 &0.791 & 1.180 \\
            Spelling Mistakes (IRR)& &1.002 &1.019 &1.090 & 1.083 \\
            Length of Review (IRR)& &1.000 &1.000 &1.001 & 1.005 \\
            Likelihood of Fake (IRR)&&0.210&0.142&0.107&0.133\\
            \hline
            Observations& &221,602&221,601&221,601&221,601\\
            \hline
            \textit{Note:}  & \multicolumn{1}{r}{$^{*}$p$<$0.05; $^{**}$p$<$0.01; $^{***}$p$<$0.001}
            \end{tabular}
            }
            \label{stat2}
\end{table*}

\subsection{Results} 

\textbf{Reviews written by new users with fewer friends and reviews are categorized as not recommended.}
We ran a multivariant logistic regression model on all the data, with \emph{recommended} as the dependent variable and the number of \textit{friends} and \textit{reviews} as the independent variables. We also controlled for various confounding metrics for review quality such as \emph{review length (in char.)}, \emph{no. of spelling mistakes}, \emph{review sentiment (positive/ negative)}, and the likelihood of being a fake review.
We used logistic regression because our dependent variable is binary. 
Our findings show that the reviews of users with more friends and reviews are more likely to be recommended. Additionally, reviews that are less likely to be fake are more likely to be recommended. To check the robustness of our method, we created a binary variable of \emph{Likehood of Fake}, where a score of less than and equal to 0.5 was labeled as 0, and greater than 0.5 was labeled as 1. Our results were the same as those obtained using the original proportion; hence, our model is robust.
However, these findings from the whole dataset might be misleading because the distributions of the number of friends and reviews variables are skewed. Therefore, we also ran two quartile logistic regressions, one for the \emph{friends} and the other for \emph{reviews}, where we could investigate the relationships in each quartile. 
Table~\ref{stat2} shows the results. 
The results on the whole data, without quartile analysis (in the second column) still show a statistically significant positive association between the number of friends and reviews and the reviews being categorized as \emph{recommended}. 
In this paper, we report Incident Rate Ratios (IRR), which are the exponentiated coefficients of Poisson regressions. These ratios show the multiplicative effect on the expected value.
Quartile regression on users' number of friends (in Table~\ref{stat2}) reveals that a strong positive association exists between the users' number of friends and their reviews being categorized as recommended than those users who have less no. of friends, with incident rate 1.001 ($Q_3$) that of users with fewer friends (a 0.1\% increase) $p< 0.001$, and 1.000 ($Q_4$) that of users with fewer friends $p< 0.001$. 

Quartile regression on users' number of reviews (in Table~\ref{stat2}) reveals a positive association between the users' number of reviews and their reviews being categorized as recommended (IRR = 1.917, $p< 0.001$) in the first quartile, where users have less number of reviews. 
However, we found that when users' review count starts to increase, i.e., from the second to the third quartile, even though there is still a positive association, the impact decreases as the incidence rate is 1.164 times (16.4\% increase) in the second quartile and then 0.016 times (1.6\% increase). Interestingly, these results are consistent when we ran the regression model on all the data (See column second of Table~\ref{stat2}) as the higher the number of friends and reviews, the more likely they would be in the recommended section, with an increase of 0.3\% and 5.1\% for friends and reviews respectively. Hence, in summary, we found \emph{support} for our \textbf{H1} that new or less-established users with fewer friends and reviews are more likely to have their reviews in the not-recommended section, indicating possible bias against these groups of users and statistically verifying what the authors in~\cite{eslami2019user} found by doing their user study.

\section{Auditing Bias of Yelp’s Ranking System}
Yelp's Ranking System has three key stakeholders, i.e., Yelp, Businesses on Yelp, \& Yelp Users where each has a different fairness perspective. 
    \textbf{Yelp} would want to ensure that its business ranking algorithm provides an accurate ranking for businesses so that both users and businesses can benefit from the platform. 
    \textbf{Businesses on Yelp} would want Yelp to remove fake reviews from their platform and be ranked fairly compared to their competitors.  
    \textbf{Yelp Users} would want Yelp to fairly provide them with the ranking of businesses. 
These three perspectives are similar: all want a fair ranking. 
However, the main contrast comes when Yelp uses additional features that explicitly or implicitly consider businesses' location and, subsequently, demographic composition attributes for ranking. This can harm businesses that are in areas of low income or education or are in diverse neighbourhoods~\cite{zehlike2017fa}.

\subsection{Methodology \& Bias Metrics}
\textbf{Data Collection: } Since the Yelp Dataset Challenge does not provide the ranking of businesses, we developed a second data collection framework to collect business rankings in a given city. We only collected organic search results, i.e., \textit{Sponsered Businesses} were not collected. Additionally, the location was set through the form on Yelp's homepage. Hence, making sure that we only get the true ranking of businesses in a city.
% To make sure we do not manually select some cities based on their popularity and incur a bias in our analysis, 
We limited our analysis to some cities, obtaining a sample of them known for their demographic and social compositions.  
We first collated a list of cities based on the following criteria: \emph{tourism}, \emph{wealth}, \emph{education}, \emph{diversity}, and \emph{crime}. To do this, we identified reputable and reliable websites, such as HGTV, CNBC, Forbes, USA Today, and WalletHub, which provided up-to-date lists of cities based on specific criteria of our interest. For each criterion, we collated the top 10 cities.  
Then, we created a 127 by 5 matrix, with each row representing the cities and the column representing the criteria that it fulfilled. 
In some cases, one city was repeated multiple times for different criteria.
For example, one source (HGTV) listed Washington D.C. as a top tourist destination, and the other (USA Today) listed it as a wealthy city; hence, in the matrix, we marked Washington D.C. for both of these categories. 
To condense this list and find representatives, we followed \emph{The Skyline} approach~\cite{asudeh2015discovering,rahman2017efficient}. In the skyline approach, we try to identify the set of non-dominated cities with different attributes, i.e., the cities that fulfill the attributes and are not dominated by any other city for those particular attributes. 
For example, when creating the skyline for the cities, we found that Anchorage dominated Asheville and Denver in \emph{tourism}, \emph{wealth}, and \emph{crime rate}; hence, we chose Anchorage as it dominated these cities. 
When multiple cities fulfilled various criteria, such as Tourism, Diversity, etc., and were all part of the skyline, we chose the cities with the highest population count. 
For example, Chicago, Los Angeles, Louisville, and New York City all satisfy tourism and diversity criteria; however, Louisville is less populated than the other three cities, so we removed it from our list. Using this approach, we obtained ten cities shown in Table~\ref{tab:citiessky}. 
We acknowledge the limitation that analyzing all the cities would be ideal. However, considering the limited number of queries enforced by the Yelp server, we had to limit our scope to a subset of cities. Additionally, even though we did not consider all US cities, we selected a representative set based on the considered criteria.

\begin{table}[!htb]
    \centering
    \caption{List of Cities with the criteria. Note that X denotes that the city fulfilled the requirement}
    \label{tab:citiessky}
    \resizebox{0.75\columnwidth}{!}{%
    \begin{tabular}{c|c|c|c|c|c}
    \hline
       City  & Tourism & Wealth & Education& Diversity& Crime Rate \\
       \hline
        Corpus Christi & & &X&X&X \\
        Detroit& X& & X& X& \\
        Seattle&X&X&&& \\
        San Jose&X&X&X&X&\\
        San Francisco&X&X&X&&\\
        New Orleans&X&&&&X\\
        Anchorage&X&X&&&X\\
        Chicago&X&&&X&\\
        Los Angeles&X&&&X&\\
        New York City&X&&&X&\\
        \hline
    \end{tabular}
    }
\end{table}    
    \hfill
\begin{table}[!htb]
\centering
    \caption{Number of Restaurants per City}
    \resizebox{0.6\columnwidth}{!}{%
    \begin{tabular}{l|c|l|c}
    \hline
     \textbf{City}&\textbf{\# Rests.} & \textbf{City}&\textbf{\# Rests.} \\
     \hline
    Anchorage & 291 & Chicago & 329\\
    Corpus Christi & 271 & Los Angeles & 344\\
    New Orleans & 281 & New York City & 348\\
    San Francisco & 353 & San Jose & 346 \\
    Seattle & 305 & & \\ \hline
    \end{tabular}
    }
    \label{rest}
\end{table}

However, we found that many restaurants from Canada appeared when searching for restaurants in Detroit. Thus, we removed Detroit from our list of cities. 
We also limited our analysis to \emph{restaurants} as the category as it is one of the most searched categories on Yelp~\cite{yelp-facts1}. 
Using our custom-built crawler, we obtained the list of all the restaurants with their corresponding ranks and addresses.
We noticed that the ranking of a restaurant is dynamic and changes at different times, even on a specific day. Therefore, we need more data points on the ranking of businesses. Based on Central Limit Theorem, we decided to collect 30 data points (i.e., rankings for a specific city) so that their means be approximately normally distributed~\cite{kwak2017central}.  
Therefore, we extracted the rankings twice a day, once at 11:00 AM and once at 7:00 PM, for 15 days, from March 4th to March 19th, 2023. 
Table~\ref{rest} shows the total number of restaurants obtained for each of the nine cities. We obtained the most number of restaurants for the city of San Francisco (353) and the least for New Orleans (281). Please note that Yelp provides a certain number of pages for each city, and we extracted all the restaurants from those pages, and also popular fast food restaurants such as Wendy's, McDonald's, etc., are often not reflected in Yelp's ranking. Since these restaurants are not in the ranking and are not shown to the users, this does not affect our results. 

During the data collection, for each restaurant, we also obtained the zip code, where the restaurant is located, and their star rating. We could not obtain the address or the latitude and longitude of these restaurants from Yelp; hence, we used the Google Places API~\cite{places-api} to extract this information. For each city, we provided the name of the restaurant as well as the latitude and longitude of the city. 
Moreover, using the zip code, we extracted the \emph{demographic}, \emph{educational}, \emph{income}, and \emph{employment} data from the US Census website of 2020~\cite{census}. The Census provides the demographic information for each zip code \url{https://tinyurl.com/2w4sbe2k}.

\textbf{Businesses' exposure by ranking: } 
To identify the disparate impact in the ranking of businesses, we computed the average exposure of each restaurant. We employed the exposure-based fairness metric proposed in~\cite{singh2018fairness}. Exposure-based fairness is defined by quantifying the expected attention received by a candidate or a group of candidates, typically by comparing their average \emph{position bias} to that of the other candidates or groups~\cite{zehlike2022fairness,joachims2017accurately}. This metric allowed us to find the exposure of the businesses to their ranking. Higher exposure means the businesses are better ranked, appear earlier in the list, and have a higher chance of being visited/ checked by them~\cite{o2006modeling}.
The standard exposure drop-off, i.e., the position bias, is given by $\frac{1}{log(1+j)}$, where \emph{j} is the potion in the ranking. If a restaurant is consistently ranked as \#1, then the exposure for each day would be 1, and if the restaurant ranking changes per day then the exposure value is related to its ranking which is between 0 to 1. The closer the value is to 1, more often the restaurant is closer to the top. 
We used this metric because if a business has a higher exposure, that means that it would be displayed on the earlier pages of the search results and probably get more consumers because of that. 
We identified some restaurants being present only for one or a few days and not throughout the whole 15 days. 
Therefore, we assigned 0 exposure on those days and averaged all their data points. Please note that we used the Central Limit Theorem to collect 30 data points, 2 data points per day, at different times to obtain a normal distribution, hence there is no major discontinuity with the exposure of restaurants that are not in top-15. 

\textbf{Hotspot Identification:}
``City hotspots'' typically refer to popular or noteworthy locations within a city that attract a lot of visitors or attention. These spots can include landmarks, tourist attractions, cultural sites, shopping districts, dining areas, entertainment venues, and vibrant neighborhoods. Restaurants are a key component of popular areas, therefore, to identify hotspots in a city, we tried to detect clusters of restaurants in close proximity. 

Using the latitude and longitude information obtained using the Google Places API, we generated various maps using the Folium library in Python. We used the DBSCAN algorithm to perform clustering~\cite{ester1996density}  on spatial data to identify city hotspots. In DBSCAN, the user defines the appropriate epsilon ($\epsilon$) value and threshold for the number of neighbors, i.e., \emph{minPts} was set by converting 1KM into radians. 
We tried multiple epsilon values, including 100 m, 200 m, 500 m, 1000 m, etc., manually cross-checked with their corresponding plot, and found that 200 m was better than others in the number of distinct clusters obtained. 
When increasing the epsilon, we only obtained one or two clusters with large cluster sizes. We provide the anonymous link to the interactive map of clusters \url{https://tinyurl.com/4wemktd2}.

\begin{figure}
    \centering
    \subfloat[Chicago]{\includegraphics[width=0.4\columnwidth]{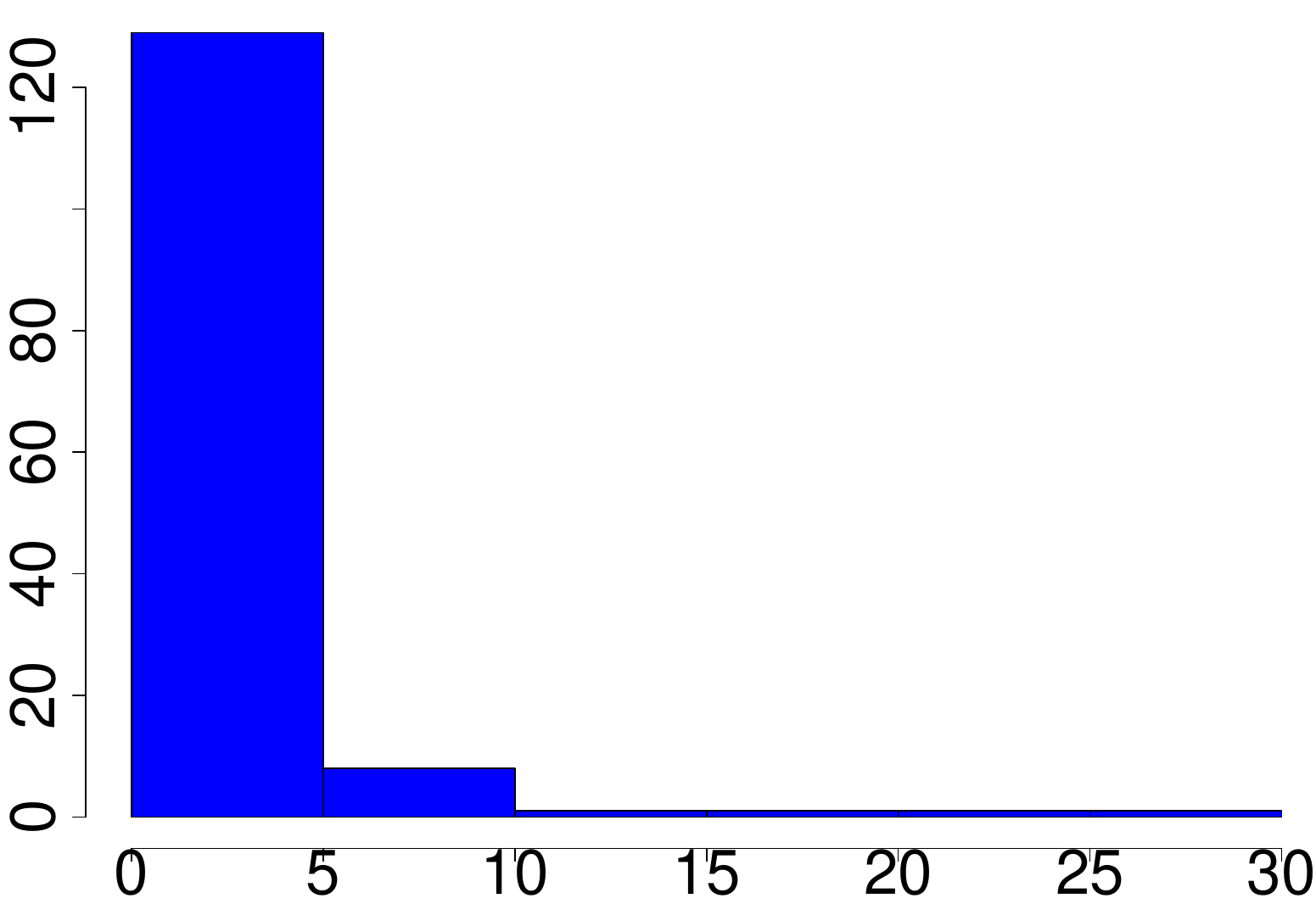} \label{fig:chig-clus}}
    \subfloat[New Orleans]{\includegraphics[width=0.4\columnwidth]{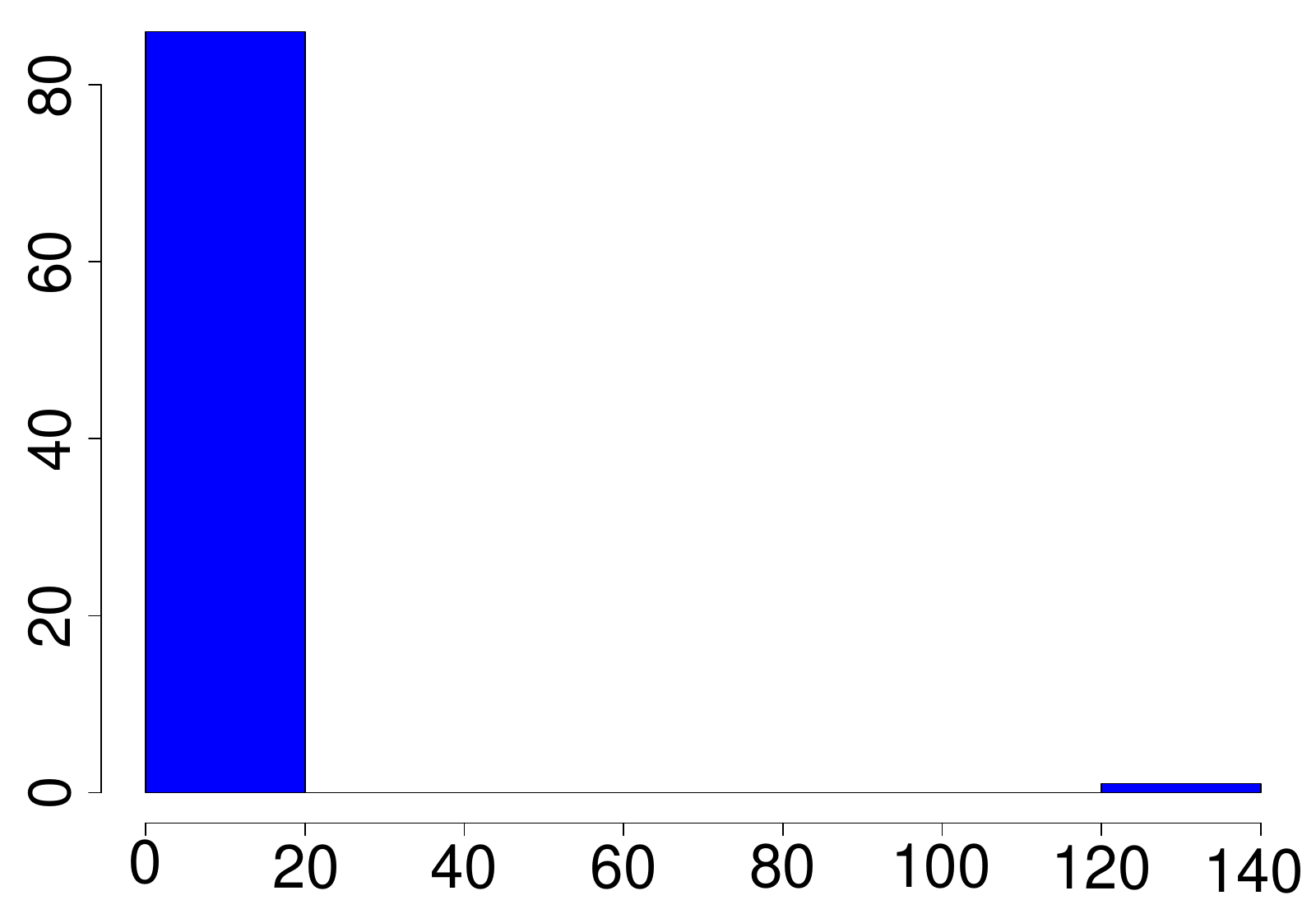} \label{fig:nola-clus}} 
    \caption{Histogram of clusters. The x-axis shows the cluster size and Y-axis shows the frequency. } 
    \label{sizecluster}
\end{figure}

We then obtained various clusters for each city containing latitude and longitude information. While in all the cities, many clusters have a size of one, we could obtain hotspots where several restaurants are located close to each other. Figure~\ref{sizecluster} shows the histogram for Chicago and New Orleans. We found that many clusters were of size 1 or 2 and not hotspots. 
We called a region a hotspot if more than 6 restaurants were in the vicinity. We chose 6 because we found that not all the cities (8) had clusters of size 7, and while only one city did not have a cluster of size 5, including them would marginally increase the number of clusters for each city. 
Hence, for each city, we created a binary variable, where we labeled a restaurant to be located in a hotspot as `1' if they were in a cluster of size six or more, else we labeled that as `0'. 
We obtained the highest number of clusters for San Francisco (13), followed by Chiacgo (12), Seattle (11), San Jose (10), New York City (9), Los Angeles (8), Anchorage (8), Corpus Christi (7), and the lowest for New Orleans (3). 
Additionally, we manually checked the obtained hotspots and found that the restaurants in those hotspots were very close, densely populated, and are in the vicinity of popular attractions in the city (for example Figure~\ref{SJC-maps} and Figure~\ref{exposure}).

\subsection{Results of Examining H2}
We statistically investigated \textit{H2} by using linear regression. 
We employed this method because our dependent variable, \textit{average exposure}, follows a normal distribution. Additionally, we included the \textit{star} rating of the business as a control variable, based on prior research that demonstrated a one-point increase in star rating leads to improved business ranking and consequently higher revenue~\cite{luca2016reviews}. We also controlled for the number of reviews and available amenities that the restaurant offers as these can also be an indication about how popular the restaurant is.
\begin{table}[!h] 
\centering 
\caption{Regression analysis to examine the association between resturants' average exposure and hotspots} 
  \label{log-yelp-attrac} 
   \resizebox{0.6\columnwidth}{!}{%
\begin{tabular}{@{\extracolsep{5pt}}ll}
\hline 
\multicolumn{2}{c}{\textit{Dependent variable: average exposure}} \\ 
\hline \\[-1.8ex] 
Hotspot& 0.023 (0.000)$^{***}$ \\ 
Stars& 0.010 (0.003)$^{**}$ \\
No. of Reviews&0.000 (0.000)$^{***}$ \\ 
Amenities&0.000 (0.001)$^{**}$\\
  \hline \\[-1.8ex] 
Observations & 2,868  \\ 
Adjusted R-squared & 0.0784 \\
\hline 
\multicolumn{2}{l}{\textit{Note: $^{*}$p$<$0.05; $^{**}$p$<$0.01; $^{***}$p$<$0.001}}
\end{tabular}
}
\end{table}

\textbf{Businesses located in hotspots have higher exposure. }
Table~\ref{log-yelp-attrac} shows the result of our regression analysis for hotspots and average exposure, when using aggregated data from all the cities. We see that there is a statistically significant positive association between restaurants' exposure, their location in hotspots, high star ratings, high number of reviews, and amenities count hence supporting our hypothesis \textit{H2}. 
We ran the same statistical test per city for all nine cities to investigate if the same pattern is seen everywhere. 
Interestingly, we observed the same results: restaurants located in hotspots are more likely to have higher exposures, hence corroborating the findings in~\cite{kokkodis2020your,huang2016effects}.
Hence, for each city, we could support our hypothesis \textit{H2}. A complete overview of the results can be found in our Appendix~\ref{appen}.  

\subsection{Results for Examining H3} 
We used logistic regression models to examine \textbf{H3} and if any relationship exists between the hotspots and demographic composition. 
Since every city has a unique structure and composition, we run the model for each city separately. 
While we analyzed all the cities individually (Appendix~\ref{appen}), due to brevity, we only present the results of Chicago and San Jose as they provided us with different findings.

\textbf{Statistical Tests:} 
We created binary variables to account for the demographic composition of each zip code in the city, namely White neighborhood (WN), Black neighborhood (BN), American-Indian neighborhood (AIN), Asian neighborhood (AN), highly educated neighborhood (HED), high unemployment neighborhood (HUne), and highly wealthy neighborhood (HWe). This was done due to in most cities, across various zip codes, the percentages of the demographic variables are close to each other, which makes it hard for the model to distinguish between them. This is a common technique used for when the relationship between a continuous predictor and the outcome is not linear.
To obtain the binary values, we used the official U.S. Census dataset quickfacts website~\cite{census-web}. We labeled `1' if the percentage of the variable is above what is given on the quickfacts website and `0' if it is not. The U.S. Census Quickfacts website gives the percentage of the white population in the city. Hence, rather than having the harsh threshold of 75.5\% which represents the overall percentage of the white population in the US~\cite{census1}, we used the percentage of the population of that city to give a more nuanced analysis. This is especially because we run the analysis on each city, and compare the demographic composition of zip codes in that city.  
For example, WN in the city of Chicago was labeled as `1' if the percentage of the White population in the zip code was greater than or equal to 42.4\%, which is the total percentage of White population in the city of Chicago. 
Whereas, WN in the city of San Jose was labeled as `1' if the percentage of the White population in the zip code was greater than or equal to 32\%, which is the total percentage of White population in the city of San Jose.
We removed the Native Hawaiian (NHN) variable from the models because the sum of all of them would be 100\%, making the variables not independent. 

We employed clustered multivariate regression clustering by the zip code. This technique combines multivariate regression analysis with clustering methods to account for intra-cluster dependencies or correlations within the data. This approach is useful when dealing with data that exhibit clustering or grouping effects, where observations within the same cluster are likely to be more similar to each other than to observations in other clusters~\cite{price2018cluster}. Here, restaurants located within the same zip code share the demographic composition characteristic of that particular zip code, and therefore, using clustering by zip code helps to account for the dependencies within the data.

\textbf{Chicago: }
We obtained 12 hotspots for Chicago. 
\begin{table}[t] 
\caption{Results of the regression analysis of hotspot with other sensitive attributes for Chicago} 
\centering
  \label{log-yelp-chica-log} 
   \resizebox{\columnwidth}{!}{%
\begin{tabular}{@{\extracolsep{5pt}}ll}
\hline 
\multicolumn{2}{c}{\textit{Dependent variable: Hotspot}} \\ 
\hline \\[-1.8ex] 
White Neighborhood (WN)& -16.485 (0.000)$^{***}$ \\ 
Black Neighborhood (BN)& -16.287 (0.000)$^{***}$ \\
American-Indian Neighborhood (AIN)& -0.142 (0.820) \\
Asian Neighborhood (AN)& -0.457 (0.446) \\
Highly Educated Neighborhood (HED)& 14.957 (0.000)$^{***}$ \\
High Unemployment Neighborhood (HUne)& -0.600 (0.407) \\
Highly Wealthy Neighborhood (HWe)& 15.706 (0.000)$^{***}$ \\
  \hline \\[-1.8ex] 
\textit{Note:}  & \multicolumn{1}{r}{$^{*}$p$<$0.05; $^{**}$p$<$0.01; $^{***}$p$<$0.001}
\end{tabular}
}
\end{table} 
\begin{figure}[t]
\centering
\subfloat[Education] {\includegraphics[width=0.5\columnwidth]{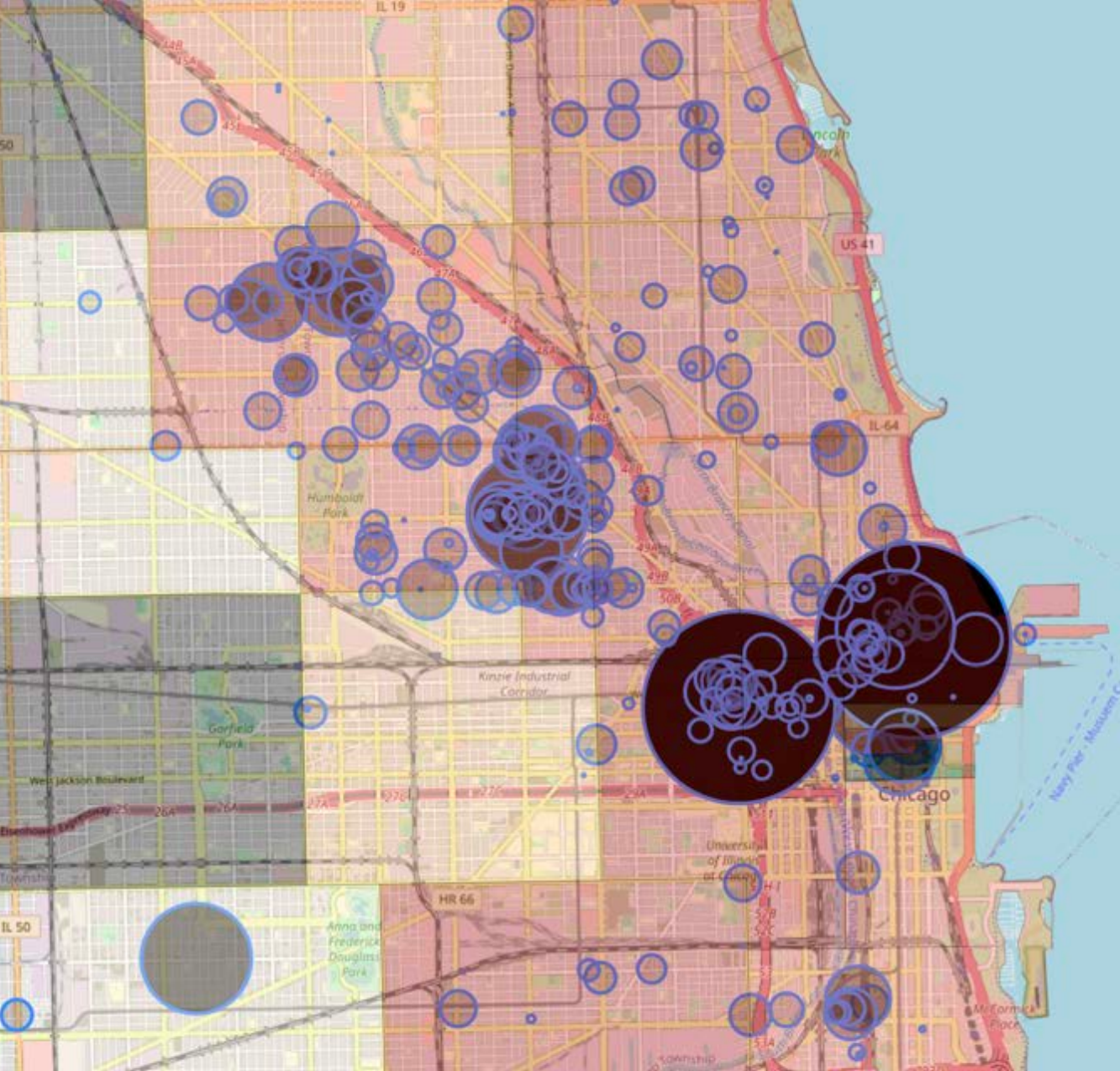} \label{fig:600}}
\subfloat[Wealth]{\includegraphics[width=0.5\columnwidth]{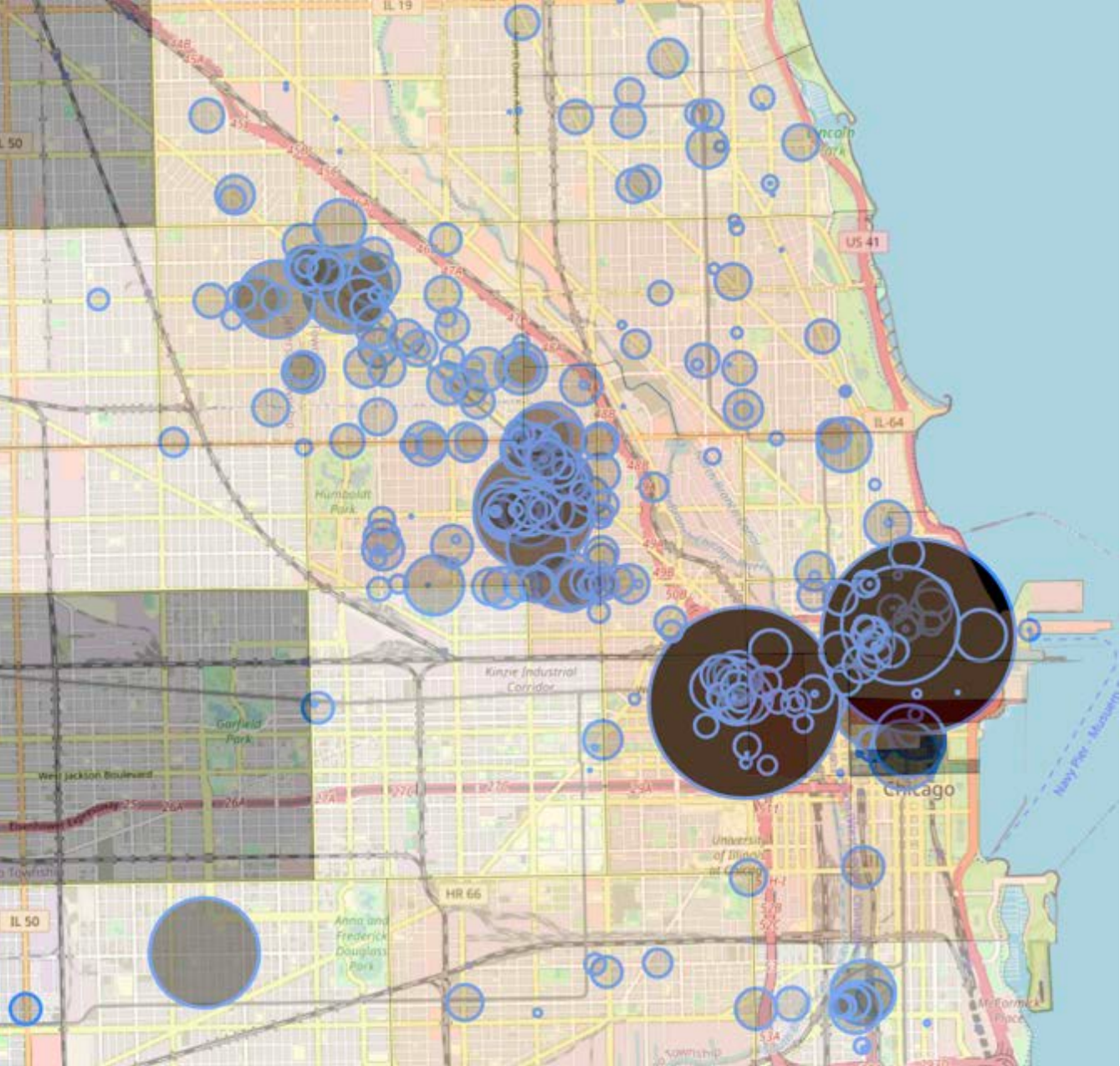} \label{fig:601}} 
\caption{Heatmap for Chicago. The black circles represent individual restaurants' exposures and colored squares represent the \% of sensitive attributes. Darker the color, higher the \% of the sensitive attribute and vice-versa.}
\label{chi-maps}
\end{figure}
Interestingly, as it is shown in Table~\ref{log-yelp-chica-log}, White and Black neighborhoods are negatively correlated to hotspots. 
We suspect this is because there are many White and Black neighborhoods that are far from the touristy parts of Chicago.
Still, when only considering the hotspot regions, we find that those are more likely to be in wealthier and higher-educated neighborhoods, hence we find support \textit{H3}. 
Our statistical results can also visually be seen in Figure~\ref{chi-maps}. We can see that major hotspots are in highly educated (shown in Fig.~\ref{fig:600}) and wealthy neighborhoods (shown in Fig.~\ref{fig:601}). 
Combining all the findings, we can conclude that businesses located in Chicago's hotspots are more likely to experience higher average exposure, particularly in wealthy and highly educated neighborhoods. This finding has significant implications for businesses situated in less affluent and less educated areas; being outside of a hotspot may result in lower exposure and, consequently, reduced revenue.

\textbf{San Jose: } 
We obtained 10 hotspots in the city of San Jose. Interestingly, we did not see a statistically significant relationship between any of the sensitive demographic attributes and the hotspots for the city of San Jose. Therefore, while we could support \textit{H2} that businesses in hotspots have higher exposure, the results do not support \textit{H3}. 
\begin{figure}[t]
\centering
\subfloat[Education] {\includegraphics[width=0.45\columnwidth]{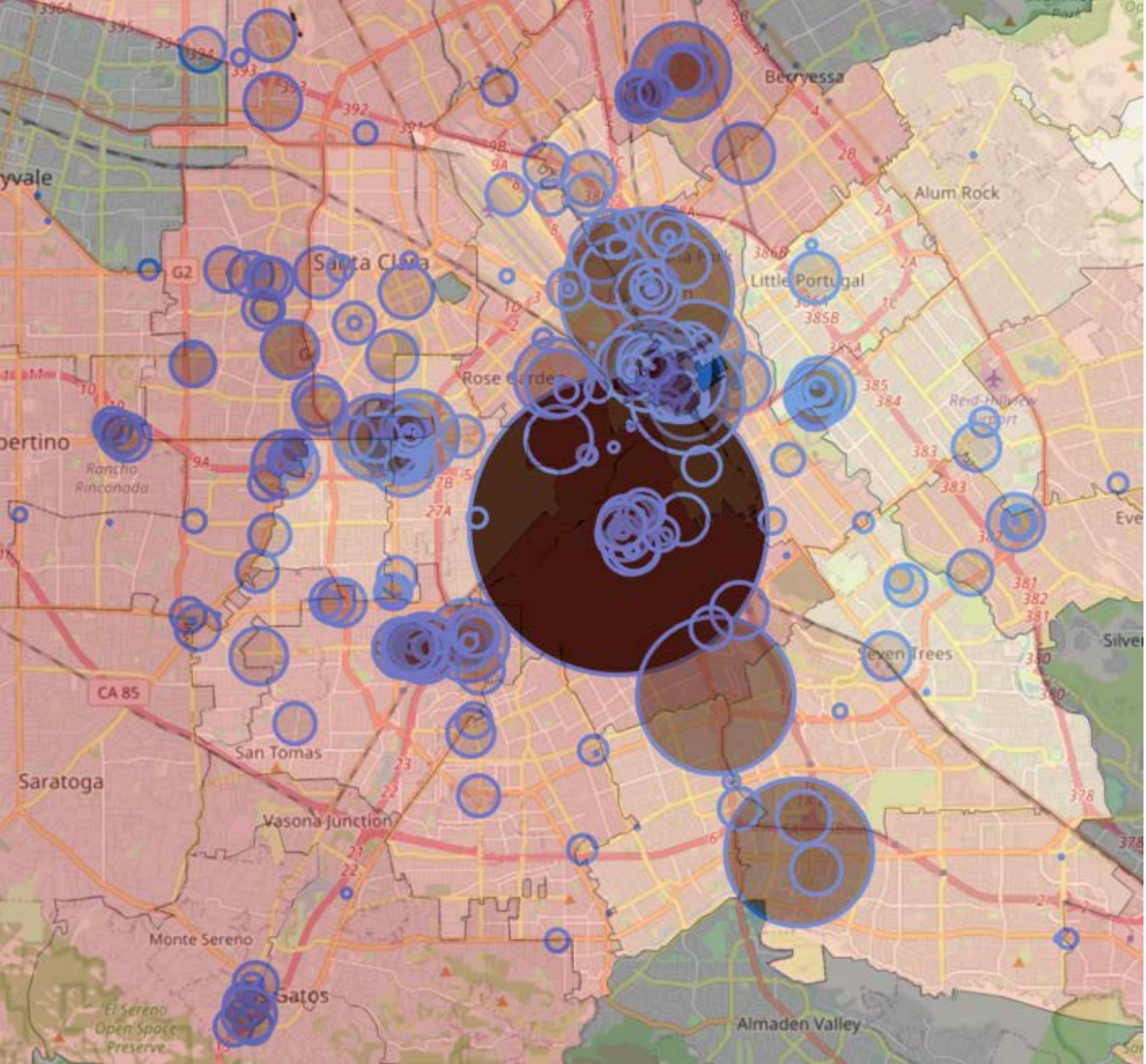} \label{fig:60781}}
\subfloat[Wealth]{\includegraphics[width=0.47\columnwidth]{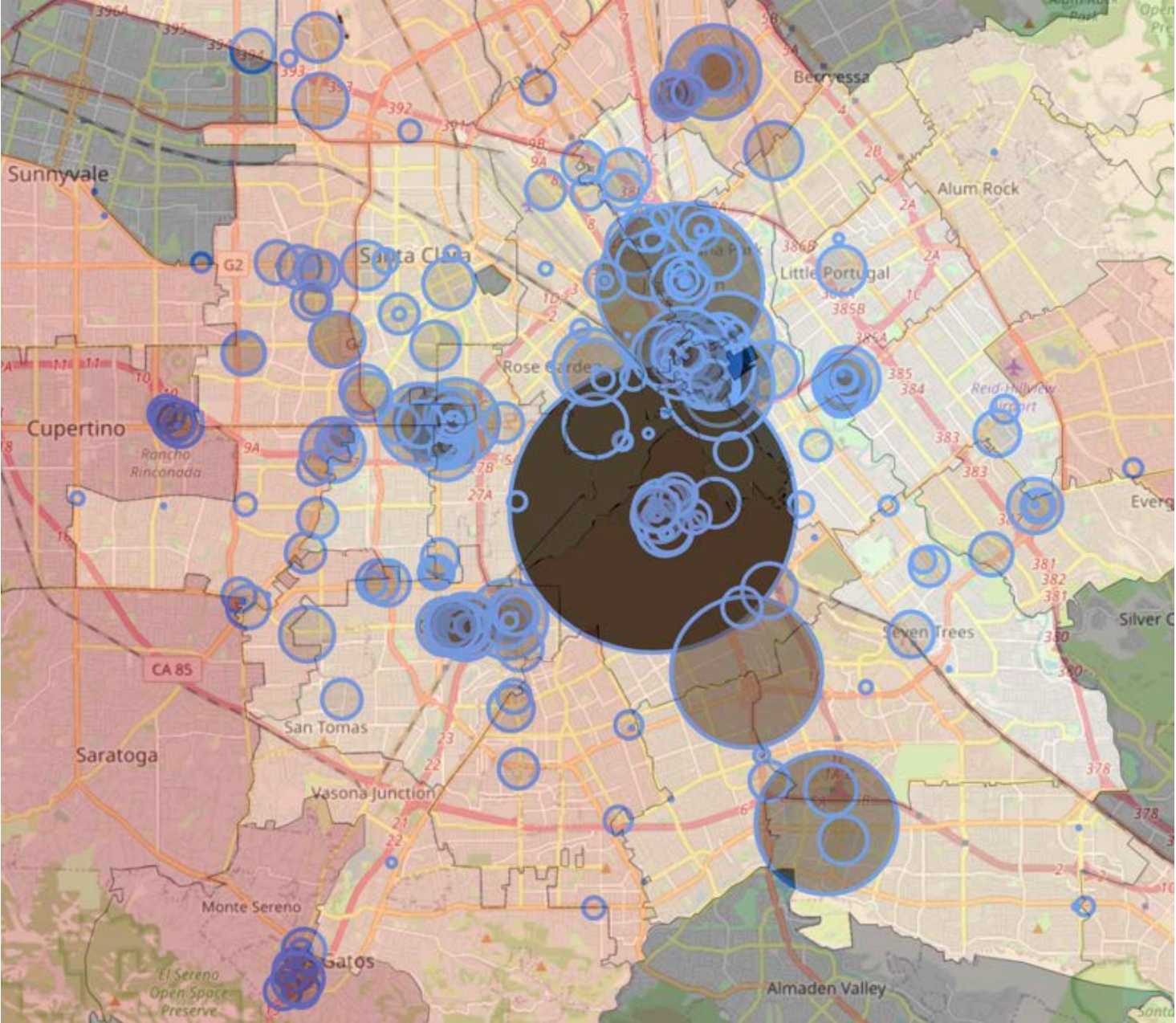} \label{fig:60w157}} 
\caption{Heatmap for San Jose}
\label{SJC-maps}

\end{figure}

This can be explained by the structure of the city as it is shown in Figure~\ref{SJC-maps}. 
This heatmap of San Jose visually shows that many regions of San Jose are characterized by higher levels of education and wealth, and hotspots are all scattered throughout the city (shown in Figure~\ref{fig:60781} and Figure.~\ref{fig:60w157}) and high-wealthy neighborhoods. 
This is in contrast to Chicago, where we observed some segregated regions in terms of wealth and education, with many hotspots being concentrated in specific areas across the city.   

\textbf{Other Cities.} 
For brevity, we do not provide the analysis for other cities. Doing the same analysis, we found support and partial support for \emph{H3} for Anchorage, New Orleans, San Francisco, and New York City. However, we did not find support for \emph{H3} for Corpus Christi, Los Angeles, and Seattle. 
We found that the cities that supported H3 tended to be more diverse, having less White population compared to the U.S. average of 75.5\%. Moreover, they
tended to have more regions with either higher education or wealth than the U.S. average of 34.3\% and \$75,149, respectively. 
These results show that the existing disparities observed in the cities are transferred as bias into Yelp's search or recommendation system, potentially creating a discriminatory feedback loop against businesses with less exposure.

\textbf{Implications:} 
Each city has unique demographic characteristics.
Hotspots appear in different regions of cities with different demographic distributions. Depending on the associations between groups and hotspots, the ranking algorithm may behave differently for those groups, which explains the implicit bias of the algorithm for different sensitive attributes within each city. 

\textbf{Broader Implication:} 
Our results highlight the importance of {\em responsible data science practices} in data-driven systems such as Yelp. Yelp and other platforms should open source their models with part of the training data for researchers to audit their systems~\cite{singhal2023sok}.

\section{Ethics and Limitation}
We only gathered reviews from businesses, and public restaurants, and did not attempt to access or find any accounts on Yelp. We also made no attempts to maliciously gather this data, instead, we used the same backend APIs that a user browser would request data from. 

\textbf{Limitations:} Firstly, we could not obtain the list of unique users. Secondly, we did not analyze the ranking of restaurants for all the cities. Furthermore, we removed Detroit from our ranking analysis. We collected the Yelp reviews in early 2023. Hence, there is a possibility that Yelp might have changed some reviews from recommended to not recommended and vice versa, and our analysis does not capture that. Additionally, our model for finding the likelihood of a review to be fake is simple, as we did not have complex features to build a sophisticated model. Furthermore, we were not able to control for the number of recent reviews and the average star rating of recent reviews because of the limit of calls we could make to Yelp and when we tried using our crawler to get them, Yelp was actively blocking us for one, two or sometimes a week and even blocking requests when using a VPN.

\section{Conclusion}
In this work, using a large-scale data-driven approach, we audit Yelp's business ranking and review recommendation software. We defined and examined three hypotheses to audit both systems. Using statistical and empirical analysis, we found that Yelp's review recommendation system disproportionately filters reviews by less-established users into the not recommended section. We also discovered that there was a linear association between a restaurant being in a hotspot and having a higher exposure. Additionally, for some cities, we found association between hotspots and some demographic composition. This can create a discriminatory feedback loop against businesses with less exposure. 

\section{Acknowledgments}
This material is based upon work supported by the National Science Foundation under Award NSF III Medium 2107296.

\bibliography{aaai25}
\section{Ethics Checklist}

\begin{enumerate}

\item For most authors...
\begin{enumerate}
    \item  Would answering this research question advance science without violating social contracts, such as violating privacy norms, perpetuating unfair profiling, exacerbating the socio-economic divide, or implying disrespect to societies or cultures?
    \textcolor{blue} {Yes}
  \item Do your main claims in the abstract and introduction accurately reflect the paper's contributions and scope?
     \textcolor{blue} {Yes}
   \item Do you clarify how the proposed methodological approach is appropriate for the claims made? 
     \textcolor{blue} {Yes}
   \item Do you clarify what are possible artifacts in the data used, given population-specific distributions?
     \textcolor{green} {NA}
  \item Did you describe the limitations of your work?
     \textcolor{blue} {Yes}
  \item Did you discuss any potential negative societal impacts of your work?
     \textcolor{blue} {No, because we ideate on positive inference from our findings. }
      \item Did you discuss any potential misuse of your work?
     \textcolor{green} {NA}
    \item Did you describe steps taken to prevent or mitigate potential negative outcomes of the research, such as data and model documentation, data anonymization, responsible release, access control, and the reproducibility of findings?
    \textcolor{blue} {Yes}
  \item Have you read the ethics review guidelines and ensured that your paper conforms to them?
    \textcolor{blue} {Yes}
\end{enumerate}

\item Additionally, if your study involves hypotheses testing...
\begin{enumerate}
  \item Did you clearly state the assumptions underlying all theoretical results?
     \textcolor{blue} {Yes, in the Hypothesis for each of the systems.}
  \item Have you provided justifications for all theoretical results?
    \textcolor{blue} {Yes, in the Hypothesis for each of the systems.}
  \item Did you discuss competing hypotheses or theories that might challenge or complement your theoretical results?
   \textcolor{blue} {Yes, in the Hypothesis for each of the systems.}
  \item Have you considered alternative mechanisms or explanations that might account for the same outcomes observed in your study?
    \textcolor{blue} {Yes, in the individual analysis Result section.}
  \item Did you address potential biases or limitations in your theoretical framework?
    \textcolor{blue} {Yes, in the Hypothesis for each of the systems.}
  \item Have you related your theoretical results to the existing literature in social science?
   \textcolor{blue} {Yes, in the Hypothesis for each of the systems.}
  \item Did you discuss the implications of your theoretical results for policy, practice, or further research in the social science domain?
    \textcolor{blue} {Yes, in the Result section for each system.}
\end{enumerate}

\item Additionally, if you are including theoretical proofs... \textcolor{green} {NA. Our study does not include theoretical proofs.}

\item Additionally, if you ran machine learning experiments... \textcolor{green} {NA. Our study does not involve any machine learning algorithm,}

\item Additionally, if you are using existing assets (e.g., code, data, models) or curating/releasing new assets...
\begin{enumerate}
  \item If your work uses existing assets, did you cite the creators?
  \textcolor{blue} {Yes}
  \item Did you mention the license of the assets?
  \textcolor{green} {NA}
  \item Did you include any new assets in the supplemental material or as a URL?
\textcolor{blue} {Yes. We include an anonymous Dropbox URL to the dataset in this submission.}
  \item Did you discuss whether and how consent was obtained from people whose data you're using/curating?
      \textcolor{green} {NA}
  \item Did you discuss whether the data you are using/curating contains personally identifiable information or offensive content?
    \textcolor{blue} {Yes}
\item If you are curating or releasing new datasets, did you discuss how you intend to make your datasets FAIR?
\textcolor{blue} {Yes, we are committed to abide by FAIR principles when sharing our dataset upon paper acceptance.}
\item If you are curating or releasing new datasets, did you create a Datasheet for the Dataset 
\textcolor{blue} {Yes, we commit to creating a Datasheet when sharing our dataset upon paper acceptance.}
\end{enumerate}

\item Additionally, if you used crowdsourcing or conducted research with human subjects... \textcolor{green} {NA. Our study does not involve research with human subjects/participants.}
\end{enumerate}

\section*{Appendix}
\subsection{Exposure of each city}

 \begin{figure*}
    \centering
    \subfloat[Anchorage]{\includegraphics[width=0.35\textwidth]{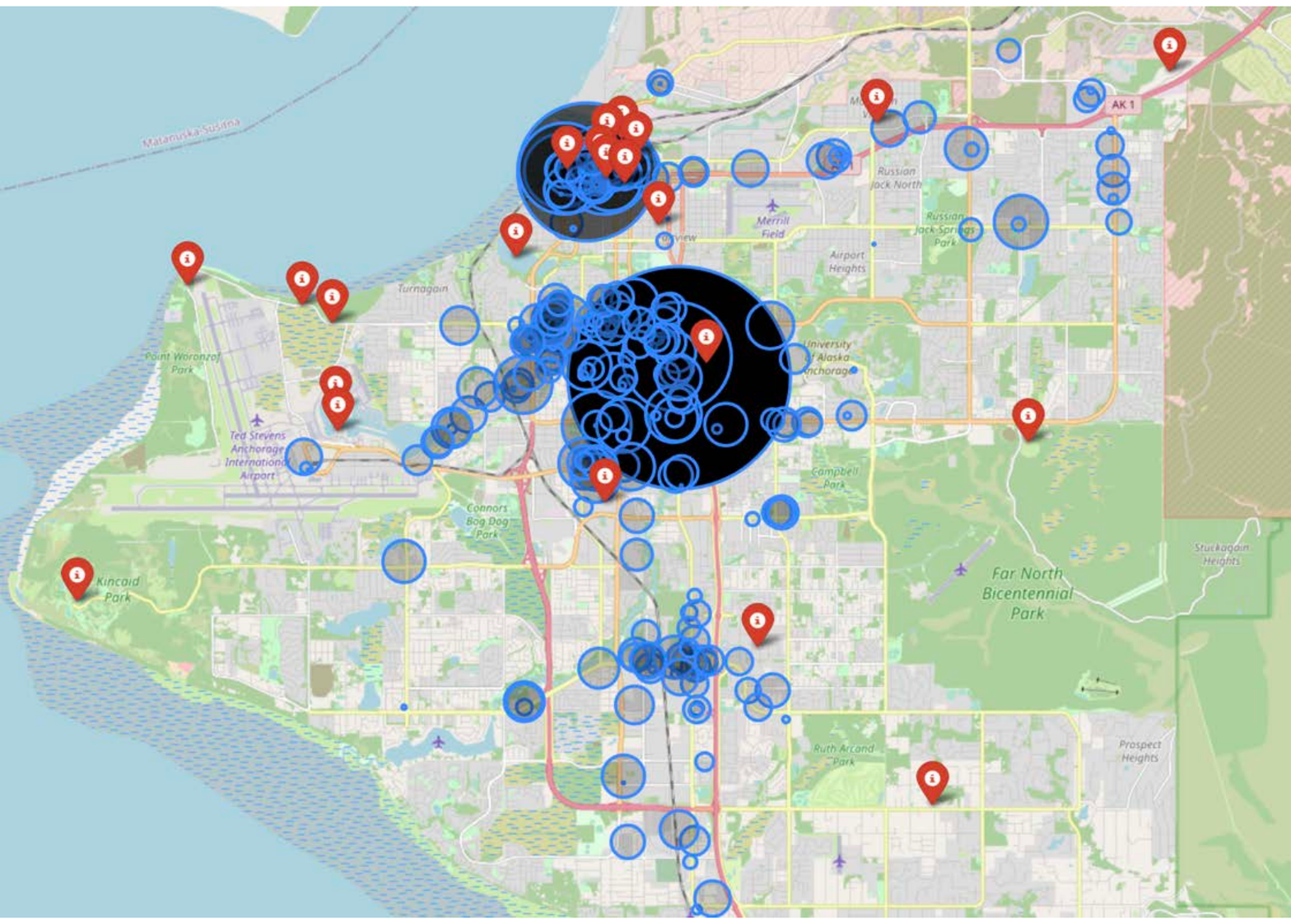} \label{fig:anch}}
    \subfloat[Chicago]{\includegraphics[width=0.25\textwidth]{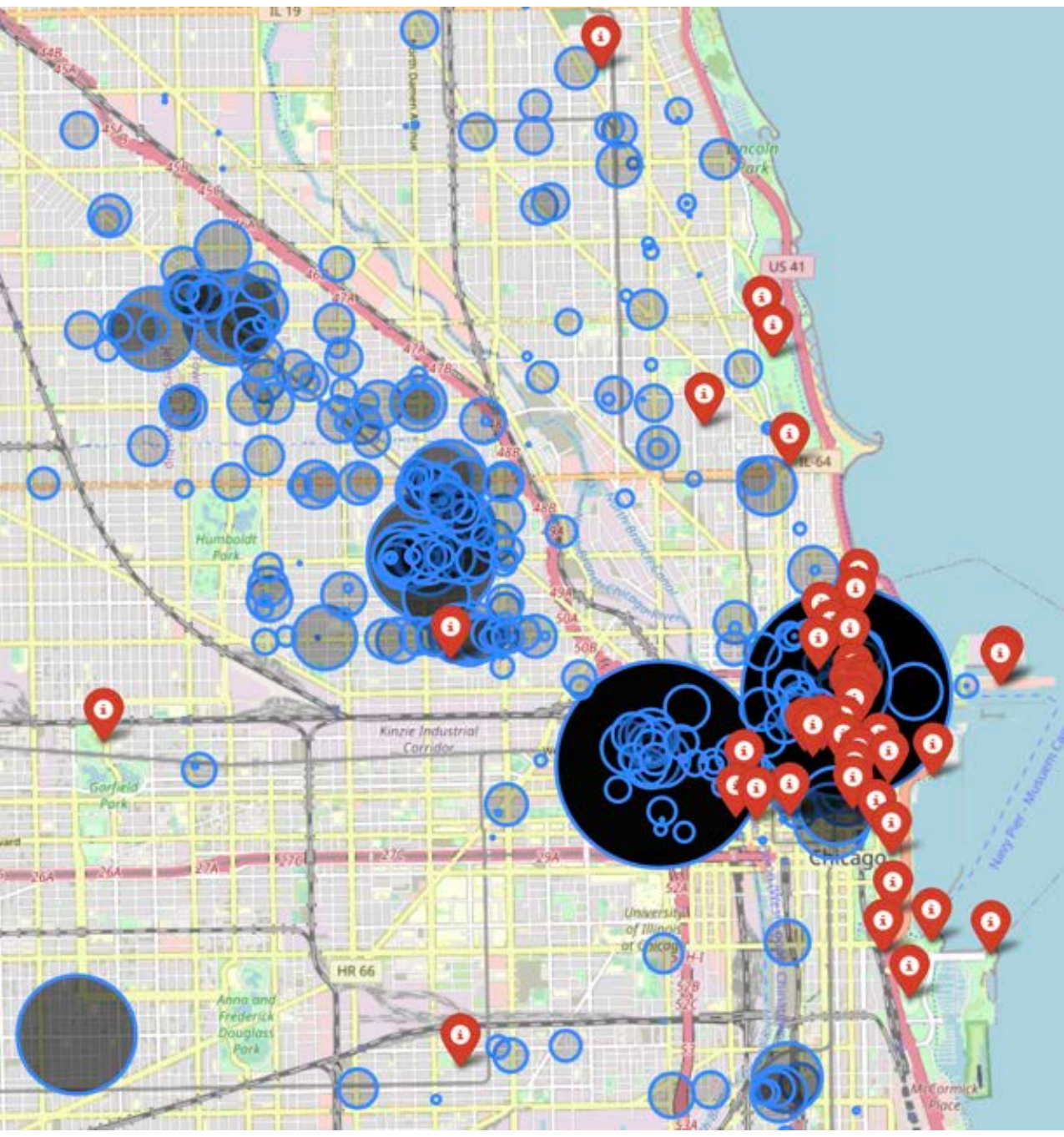} \label{fig:chig}}
    \subfloat[Corpus Christi]{\includegraphics[width=0.35\textwidth]{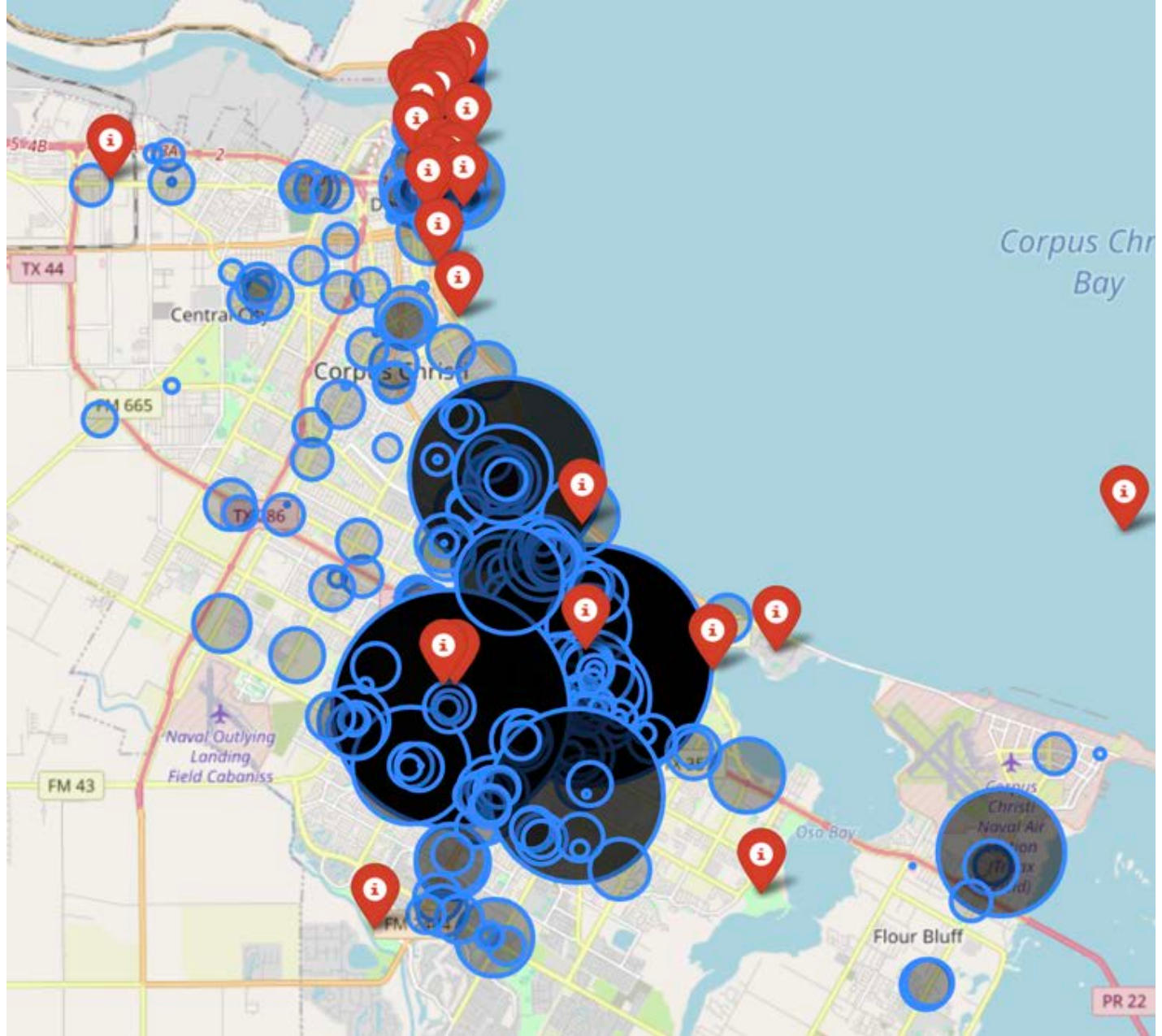} \label{fig:corpus}} \\
    \subfloat[Los Angeles] {\includegraphics[width=0.35\textwidth]{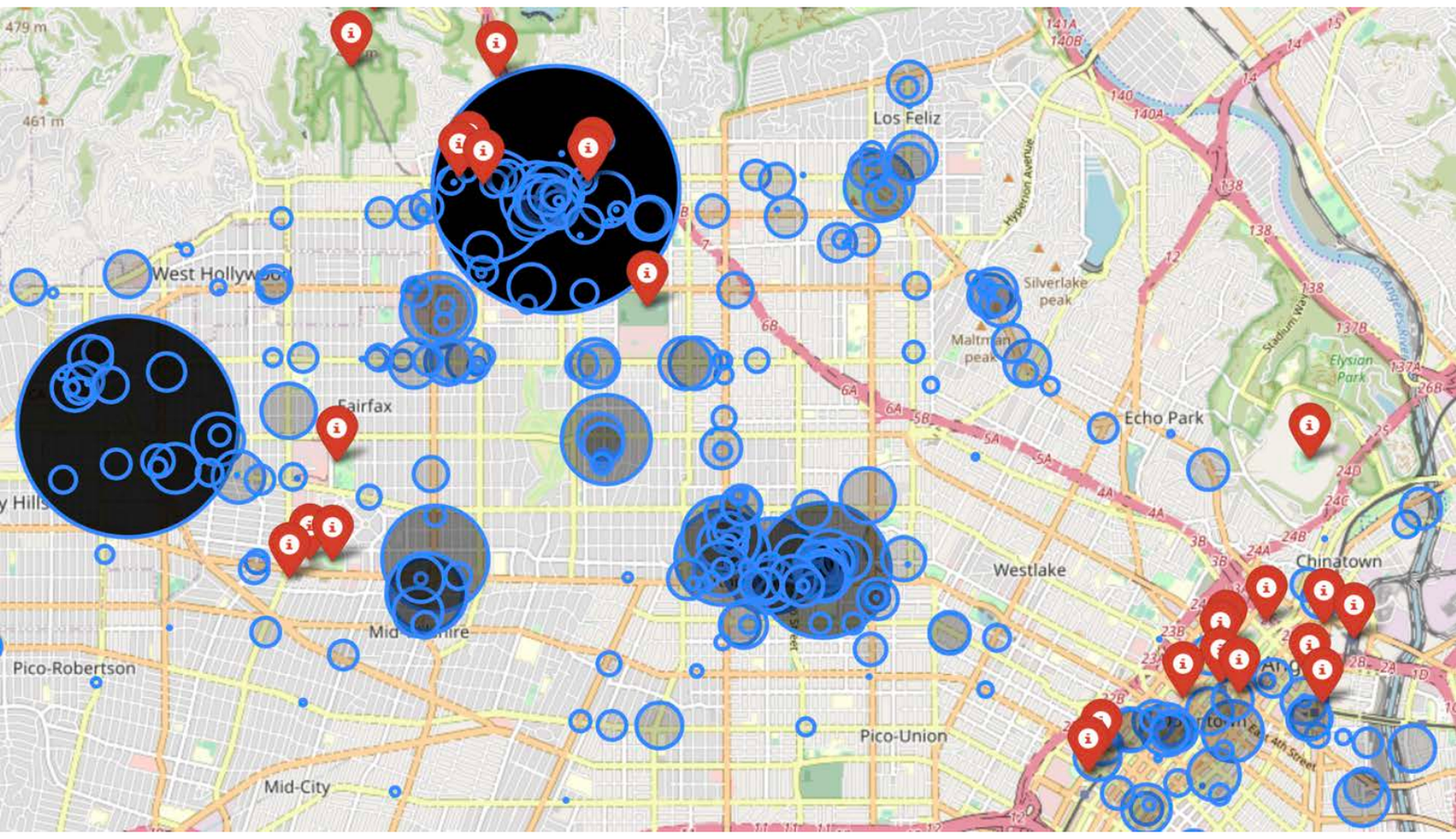} \label{fig:LA}} 
    \subfloat[New Orleans]{\includegraphics[width=0.35\textwidth]{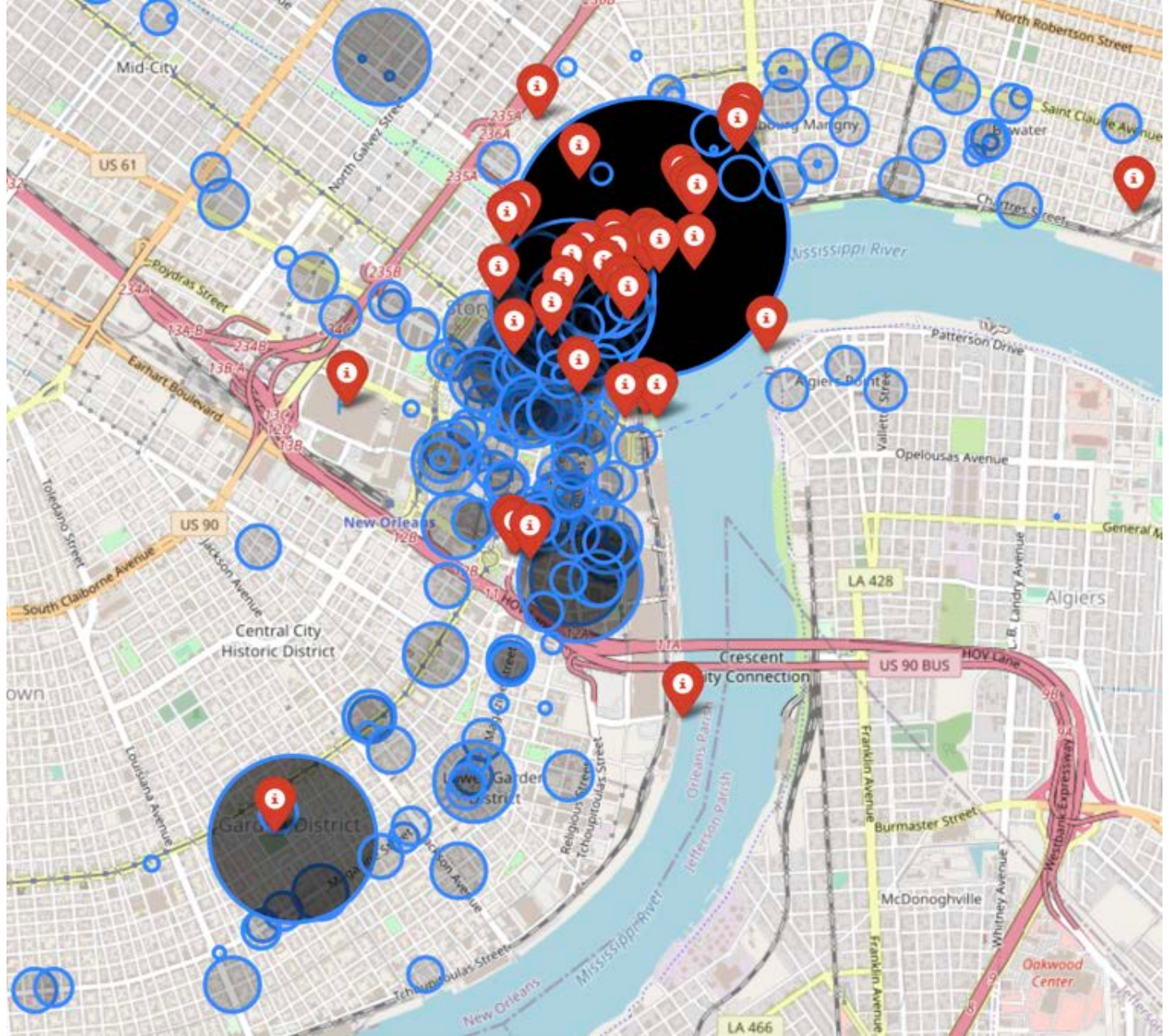} \label{fig:nola}} 
    \subfloat[New York City]{\includegraphics[width=0.25\textwidth]{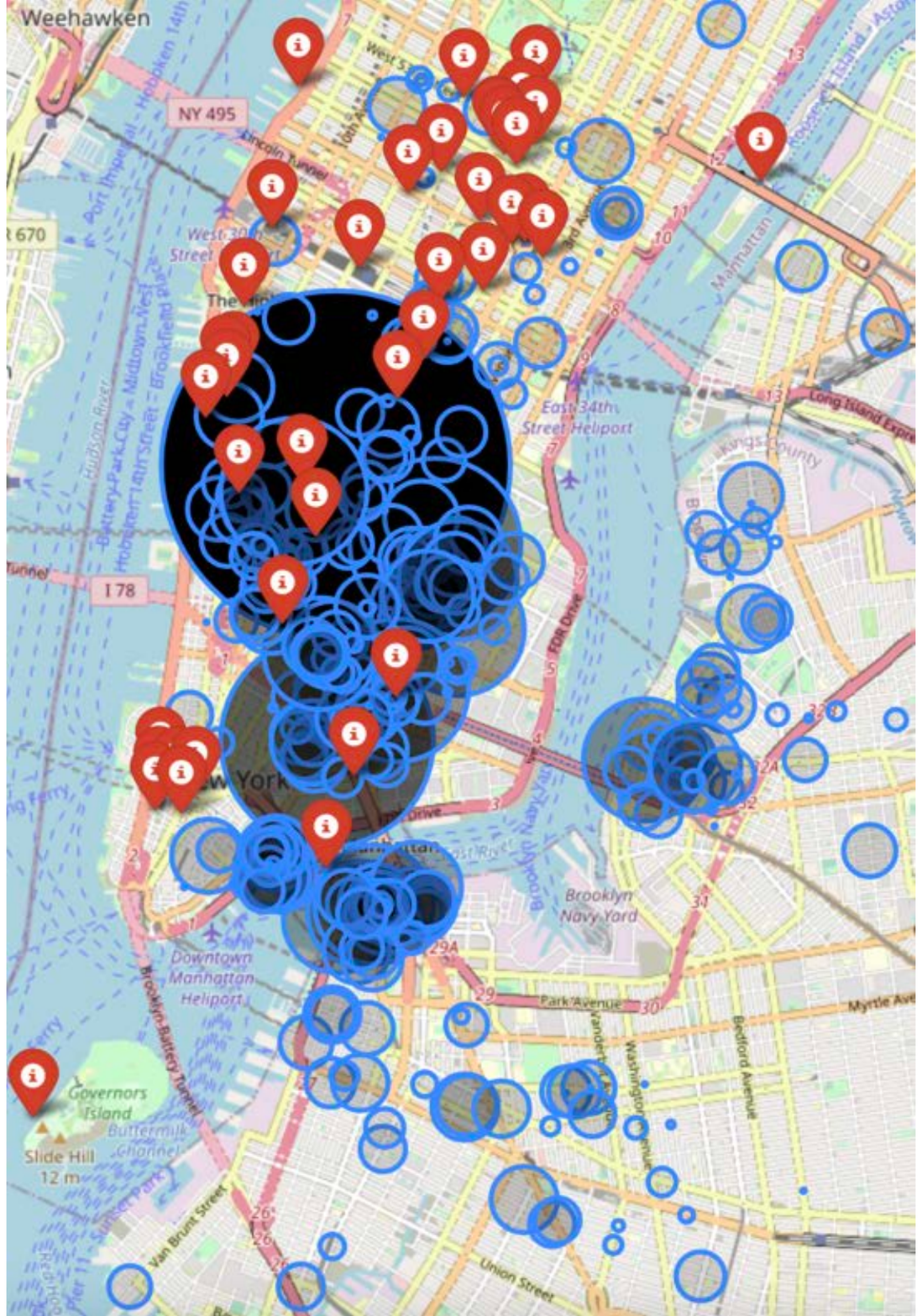} \label{fig:nyc}} \\
    \subfloat[San Francisco]
    {\includegraphics[width=0.35\textwidth]{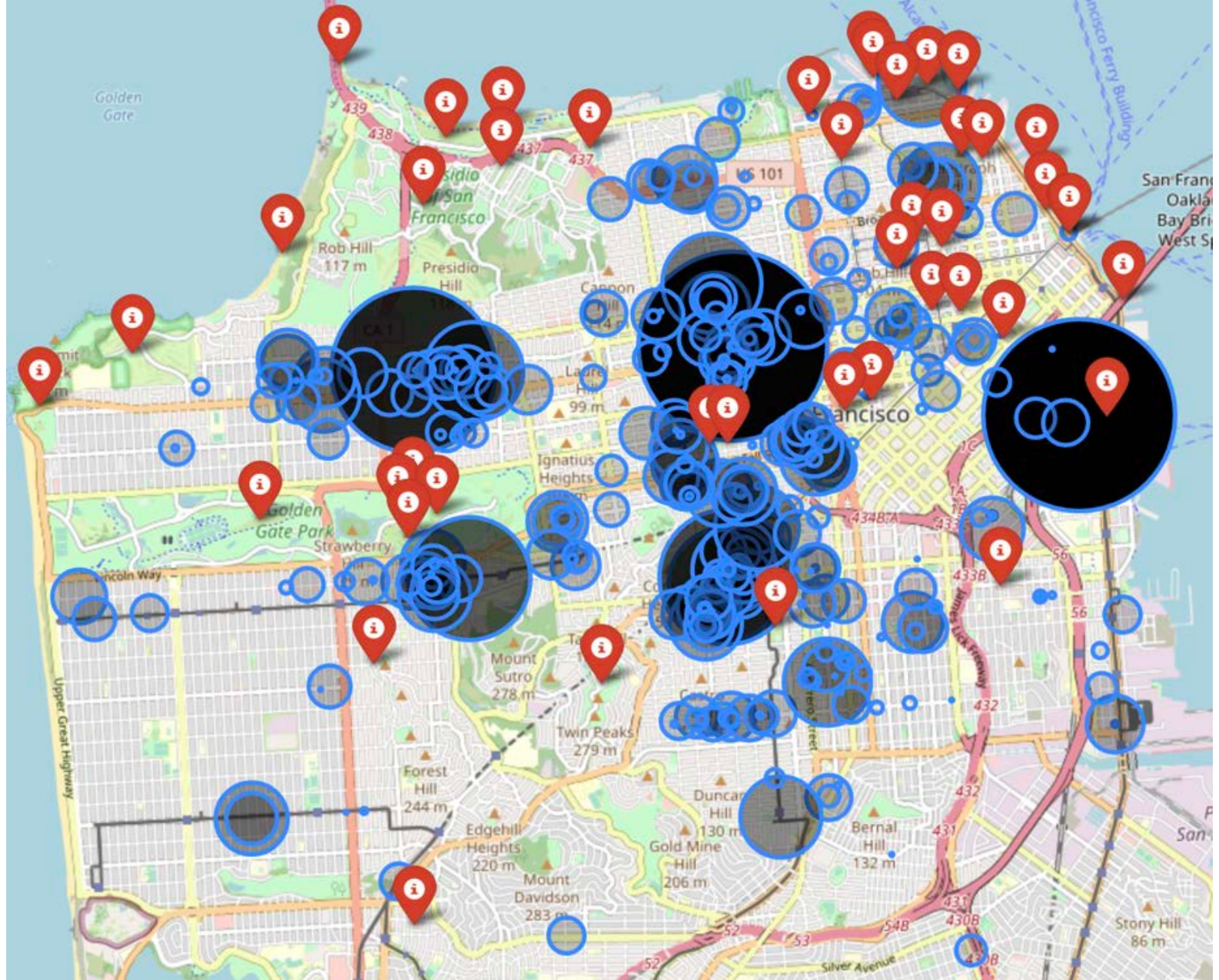} \label{fig:sanfan}} 
    \subfloat[San Jose]{\includegraphics[width=0.35\textwidth]{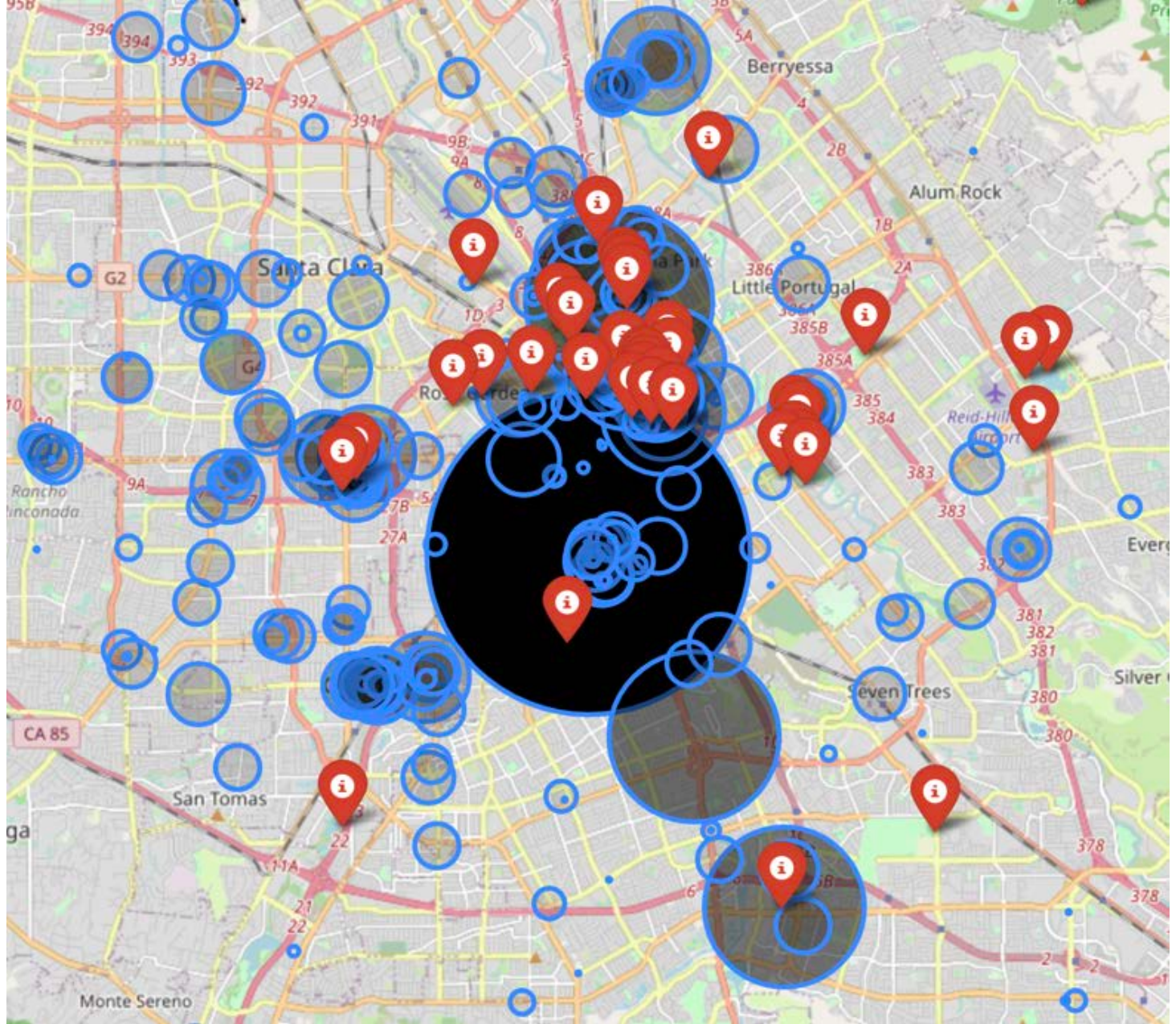} \label{fig:sanjose}}
    \subfloat[Seattle]
    {\includegraphics[width=0.25\textwidth]{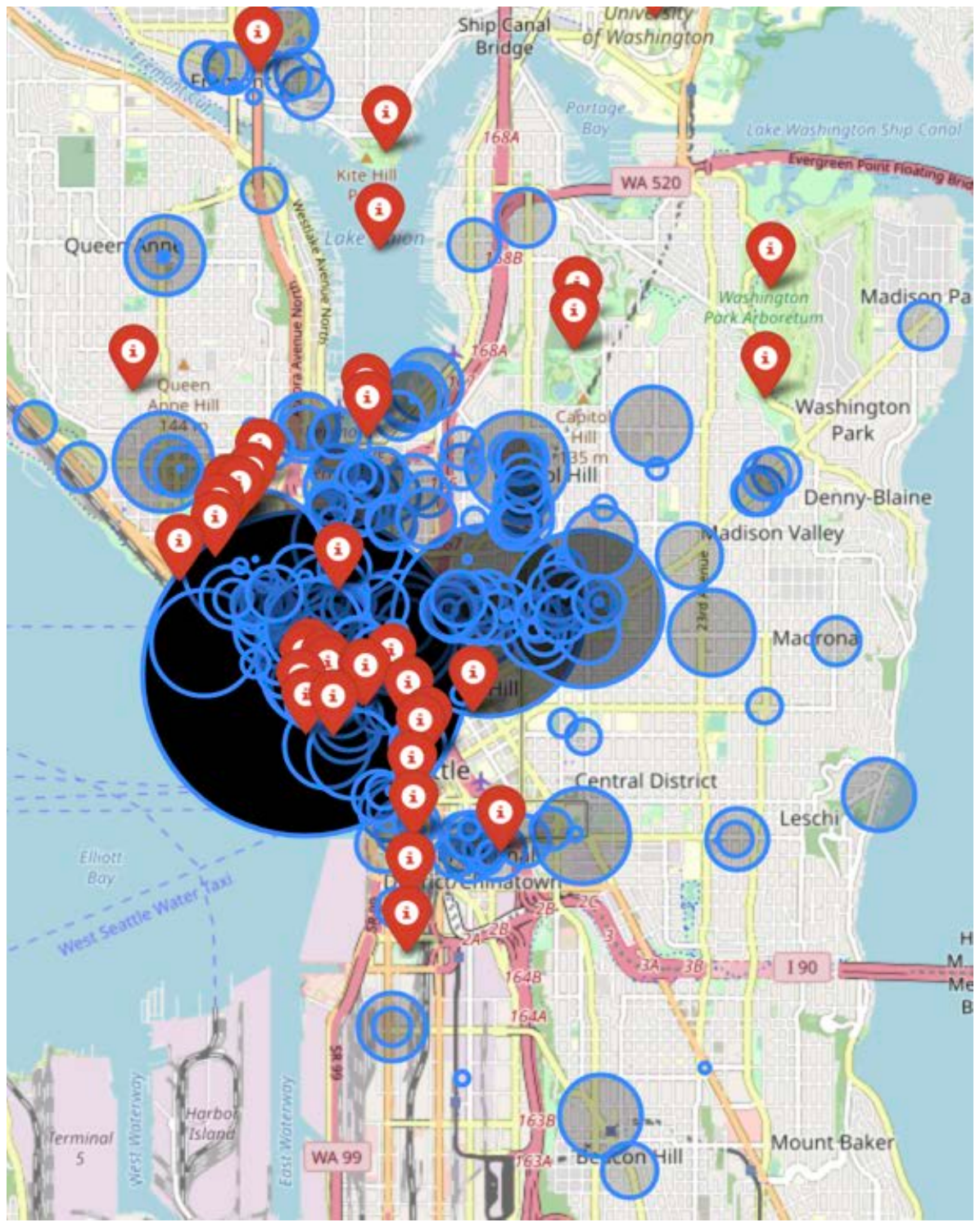} \label{fig:seat}}\\
    \caption{Exposure of restaurants in each city. Note that the red marker in the figure shows the popular attraction in each city 
    } 
    \label{exposure}
\end{figure*}

Figure~\ref{exposure} shows the exposure of restaurants per city. The circles shown in the figure show the exposure of each restaurant. The bigger the size, more is the exposure of that restaurant. 

\textbf{Anchorage: } Figure~\ref{fig:corpus} shows the exposure of restaurants in the city of Anchorage. 
By employing the DBSCAN algorithm, we obtained a total of 111 clusters. We found that there were 64 clusters of size 1. We obtained a total of 8 hotspots for the city of Anchorage. 
We then performed a linear regression to see if statistically there is a correlation between average exposure and hotspots. 

\begin{table}[h] 
\centering 
\caption{Results of the regression analysis of average exposure with hotspot for Anchronage} 
  \label{log-yelp-anchro-linearr} 
   \resizebox{0.6\columnwidth}{!}{%
\begin{tabular}{@{\extracolsep{5pt}}ll}
\hline 
& \multicolumn{1}{c}{\textit{Dependent variable:}} \\ 
\hline
 \multicolumn{2}{c}{\textit{Average Exposure}} \\ 
% \\[-1.8ex] & \textit{Linear} \\ 
\hline \\[-1.8ex] 
Hotspot& 0.033 (0.001)$^{**}$ \\ 
  \hline \\[-1.8ex] 
Observations &291  \\ 
% Multiple R-squared & 0.03412 \\
Adjusted R-squared & 0.03078 \\
% F-statistic & 10.21 \\
\hline 
\textit{Note:}  & $^{*}$p$<$0.05; \\ $^{**}$p$<$0.01; $^{***}$p$<$0.001 \\ 
\end{tabular}
}

\caption{Results of the regression analysis of hotspot with other sensitive attributes for Anchronage } 
  \label{log-yelp-anchro-log} 
   \resizebox{0.6\columnwidth}{!}{%
\begin{tabular}{@{\extracolsep{5pt}}ll}
\hline 
& \multicolumn{1}{c}{\textit{Dependent variable:}} \\ 
\hline
 \multicolumn{2}{c}{\textit{Hotspot}} \\ 
 % & \textit{Logistic} \\ 
\hline \\[-1.8ex] 
WN& -0.658 (0.003)$^{***}$ \\ 
BN& 0.512 (0.000)$^{***}$ \\
AIN& 0.972 (0.000)$^{***}$ \\
AN& -0.514 (0.000)$^{***}$ \\
HED& 0.830 (0.001)$^{***}$ \\
HUne& -1.720 (0.000)$^{***}$ \\
HWe& 0.170 (0.595) \\
  \hline \\[-1.8ex] 
\textit{Note:}  & $^{*}$p$<$0.05; \\ $^{**}$p$<$0.01; & $^{***}$p$<$0.001 \\ 
\end{tabular}
}

\end{table}

Table~\ref{log-yelp-anchro-linearr} shows the results. We observed a consistent result with that of the analysis carried out on the whole dataset, where is a positive correlation between average exposure and the restaurant being in a hotspot, hence we find support for  \textit{H2}. We then investigated if a restaurant that is in a hotspot has any relation with demographic features such as racial composition, percentage of educated people, percentage of unemployed people, and percentage of wealthy people i.e., \textit{H3}.
Table~\ref{log-yelp-anchro-log} shows the results from our regression model. We performed the analysis, by clustering them based on zip codes. We can see that restaurants that are in hotspots have a higher Black and American Indian population, they are also in highly educated and wealthy neighborhoods and hence we find partial support for  \textit{H3}

\begin{figure}[htbp!]
\centering
\subfloat[White Population] {\includegraphics[width=0.5\columnwidth]{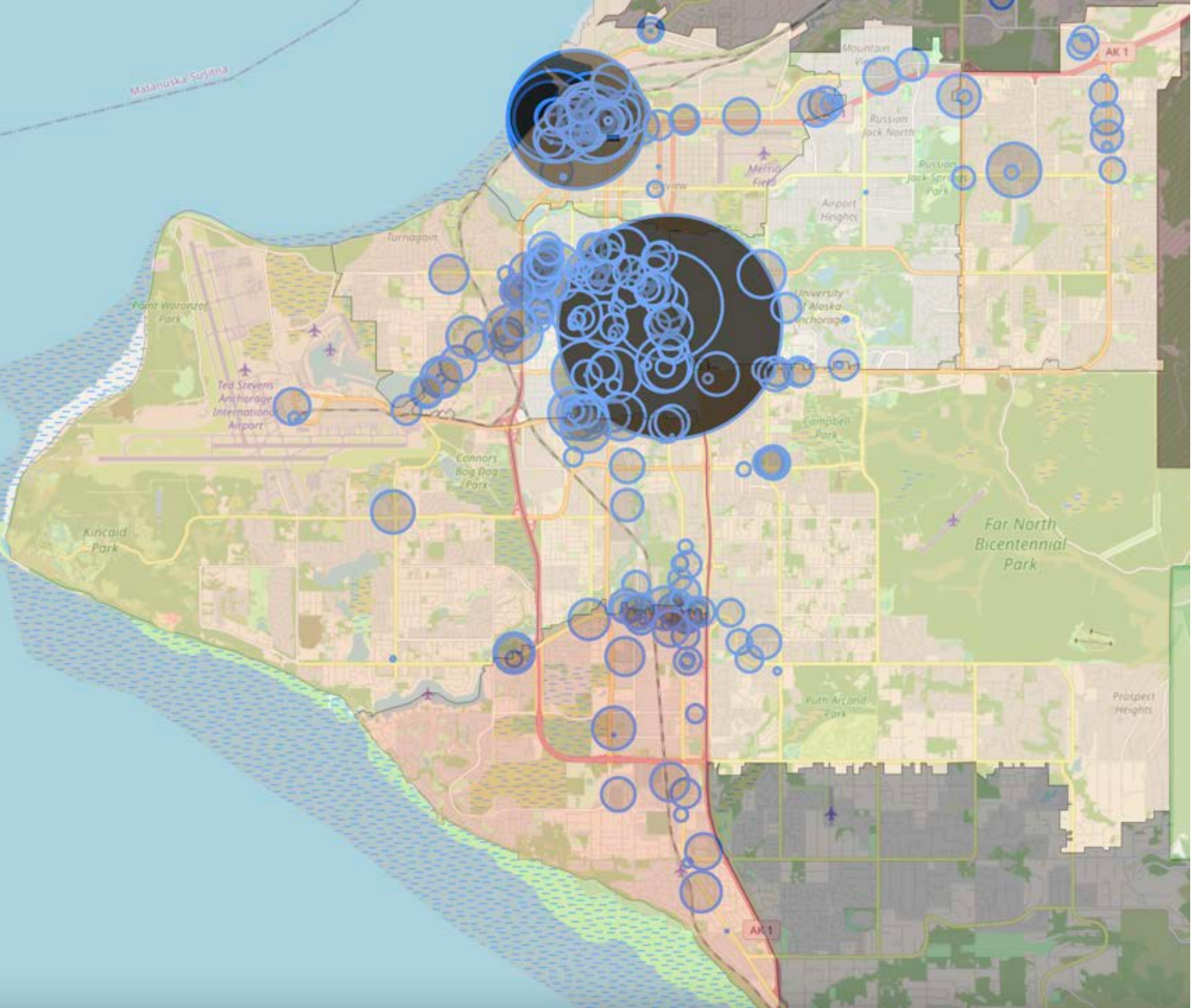} \label{fig:678}}
\subfloat[Education]{\includegraphics[width=0.5\columnwidth]{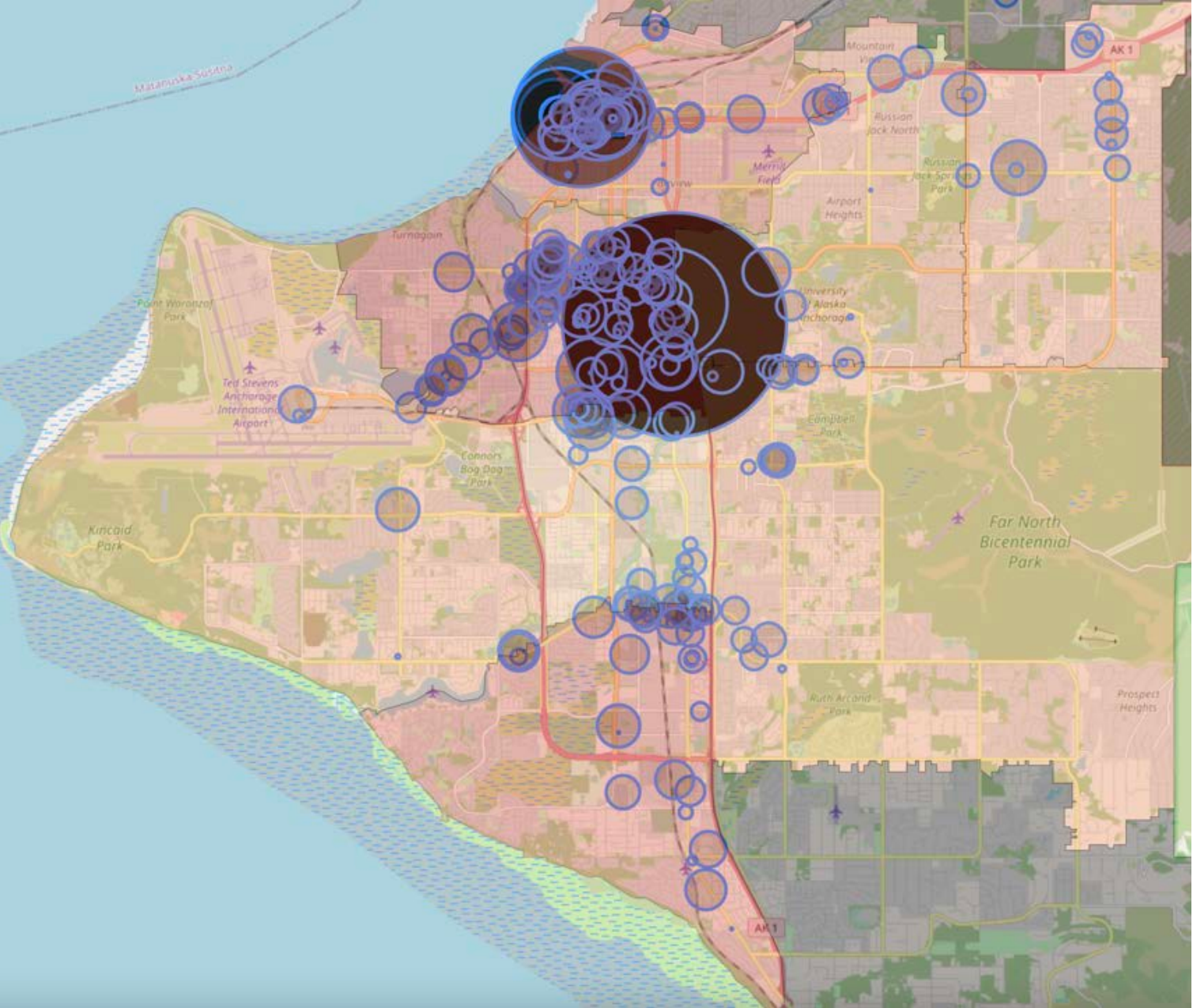} \label{fig:679}} 
\caption{Heatmap for Anchronage}
\label{anch-maps}
\end{figure}

Figure~\ref{anch-maps} shows the map for educated and white neighborhoods in Anchorage. Visually looking at the map, it confirms our statistical analysis, as restaurants that are in hotspots have higher education (shown in Fig.~\ref{fig:679}) and restaurants that are in predominantly white neighborhoods are not in hotspots (shown in Fig.~\ref{fig:678}). Therefore, we can draw a conclusion about Yelp in the city of Anchorage, businesses that are in the hotspots, have higher average exposure, and businesses that are in the zip codes with a hotspot, have a higher black and American Indian population, they are also in highly educated and wealthy neighborhoods. 

\textbf{Chicago: } Figure~\ref{fig:chig} shows the exposure of restaurants in the city of Chicago. We obtained 12 hotspots of size 6 or more. Table~\ref{log-yelp-chica-linearr} shows the results of our linear regression model. 

\begin{table}[h] 
\centering 
\caption{Results of the regression analysis of average exposure with hotspot for Chicago} 
  \label{log-yelp-chica-linearr} 
   \resizebox{0.6\columnwidth}{!}{%
\begin{tabular}{@{\extracolsep{5pt}}ll}
\hline 
& \multicolumn{1}{c}{\textit{Dependent variable:}} \\ 
\hline
\\[-1.8ex] & \multicolumn{1}{c}{\textit{average exposure}} \\ 
\\[-1.8ex] & \textit{Linear} \\ 
\hline \\[-1.8ex] 
Hotspot& 0.037 (0.000)$^{***}$ \\ 
  \hline \\[-1.8ex] 
Observations &329  \\ 
% Multiple R-squared & 0.04303 \\
Adjusted R-squared & 0.0401 \\
% F-statistic & 14.7 \\
\hline 
\textit{Note:}  & \multicolumn{1}{r}{$^{*}$p$<$0.05; $^{**}$p$<$0.01; $^{***}$p$<$0.001} \\ 
\end{tabular}
}
\end{table} 
We again see a very similar result, where there is a positive correlation between average exposure and hotspot, hence we find support for \textit{H2}.

\textbf{Corpus Christi: } Figure~\ref{fig:corpus} shows the exposure of restaurants in the city of Corpus Christi. 
Using the DBSCAN algorithm, we were able to identify 132 clusters. 
Using our threshold, we found 7 hotspots that were of size six or more. Table~\ref{log-yelp-corpus-linearr} shows the results of our linear regression model. 
\begin{table}[htbp!] 
% \begin{minipage}{0.49\columnwidth}
\centering 
\caption{Results of the regression analysis of average \\ exposure with hotspot for Corpus Christi} 
  \label{log-yelp-corpus-linearr} 
   \resizebox{0.6\columnwidth}{!}{%
\begin{tabular}{@{\extracolsep{5pt}}ll}
\hline 
& \multicolumn{1}{c}{\textit{Dependent variable:}} \\ 
\hline
\\[-1.8ex] & \multicolumn{1}{c}{\textit{average exposure}} \\ 
\\[-1.8ex] & \textit{Linear} \\ 
\hline \\[-1.8ex] 
Hotspot& 0.027 (0.024)$^{*}$ \\ 
  \hline \\[-1.8ex] 
Observations &271  \\ 
% Multiple R-squared & 0.01866 \\
Adjusted R-squared & 0.01502 \\
% F-statistic & 5.116 \\
\hline 
\textit{Note:}  & \multicolumn{1}{r}{$^{*}$p$<$0.05; $^{**}$p$<$0.01; $^{***}$p$<$0.001} \\
\end{tabular}
}
% \end{minipage}
\end{table}
\begin{table}
\centering
% \begin{minipage}{0.35\columnwidth}
\caption{Results of the regression analysis of hotspot \\ with other sensitive attributes for Corpus Christi} 
  \label{log-yelp-copus-log} 
   \resizebox{0.6\columnwidth}{!}{%
\begin{tabular}{@{\extracolsep{5pt}}ll}
\hline 
& \multicolumn{1}{c}{\textit{Dependent variable:}} \\ 
\hline
\\[-1.8ex] & \multicolumn{1}{c}{\textit{Hotspot}} \\ 
\\[-1.8ex] & \textit{Logistic} \\ 
\hline \\[-1.8ex] 
WN& 0.854 (0.457) \\ 
BN& 0.730 (0.444) \\
AIN& -15.037 (0.000)$^{***}$ \\
AN& -13.889 (0.000)$^{***}$ \\
HED& 13.668 (0.000)$^{***}$ \\
HUne& -0.241 (0.810) \\
HWe& -0.504 (0.601) \\
  \hline \\[-1.8ex] 
\textit{Note:}  & \multicolumn{1}{r}{$^{*}$p$<$0.05; $^{**}$p$<$0.01; $^{***}$p$<$0.001} \\
\end{tabular}
}
% \end{minipage}
\end{table} 
We can see that average exposure and hotspots have a positive correlation. Table~\ref{log-yelp-copus-log} shows the results of our logistic regression model. We see that restaurants in zip codes that are in a hotspot are in highly educated neighborhoods and have less number of American Indians and Asian populations. This is because Corpus Christi has a high Hispanic population.
\begin{figure}[htbp!]
    \centering
    \includegraphics[width=0.5\columnwidth]{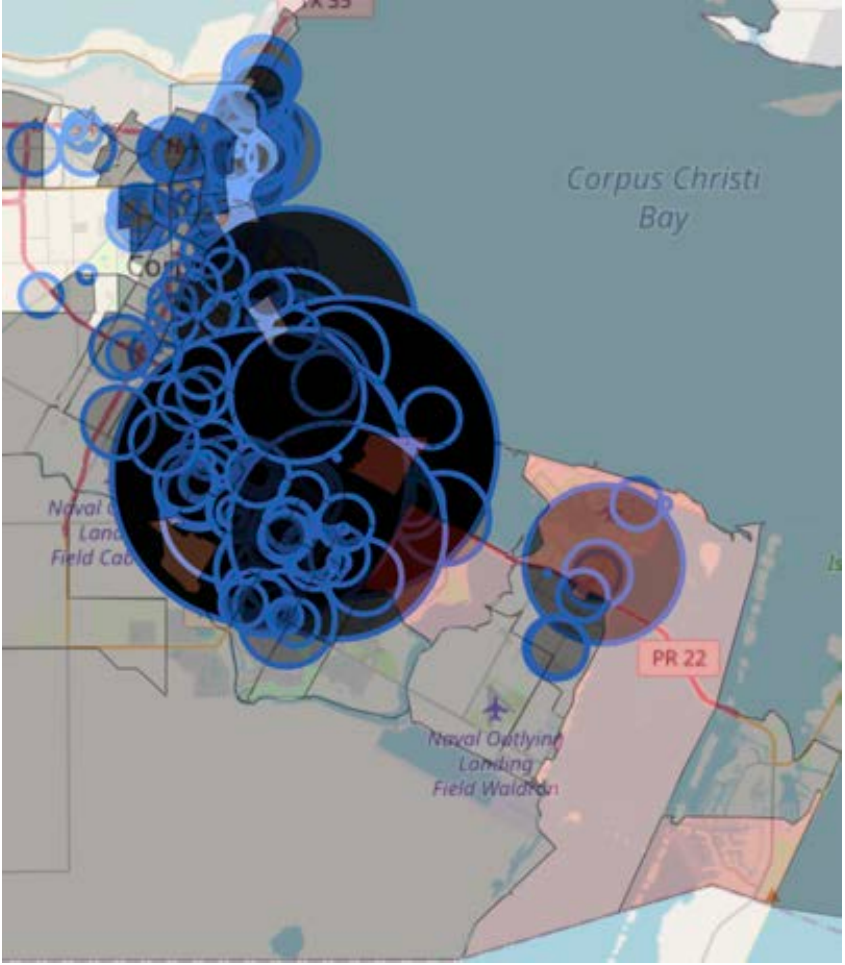} 
    \caption{Education heatmap for Corpus Christi} 
\label{figurew1}
\end{figure}
We can see the results in Figure.~\ref{figurew1}. We can visually see that hotspots are in areas that are highly educated neighborhoods. Therefore, we can draw a conclusion about Yelp in the city of Corpus Christi, businesses that are in the hotspots, have higher average exposure, and businesses that are in zip codes that are in hotspots, are in highly educated neighborhoods and have less number of American Indians and Asian population, hence, while we find support for \textit{H2}, we rejected our \textit{H3}. 

\begin{figure}[htbp!]
\centering
\subfloat[Education] {\includegraphics[width=0.5\columnwidth]{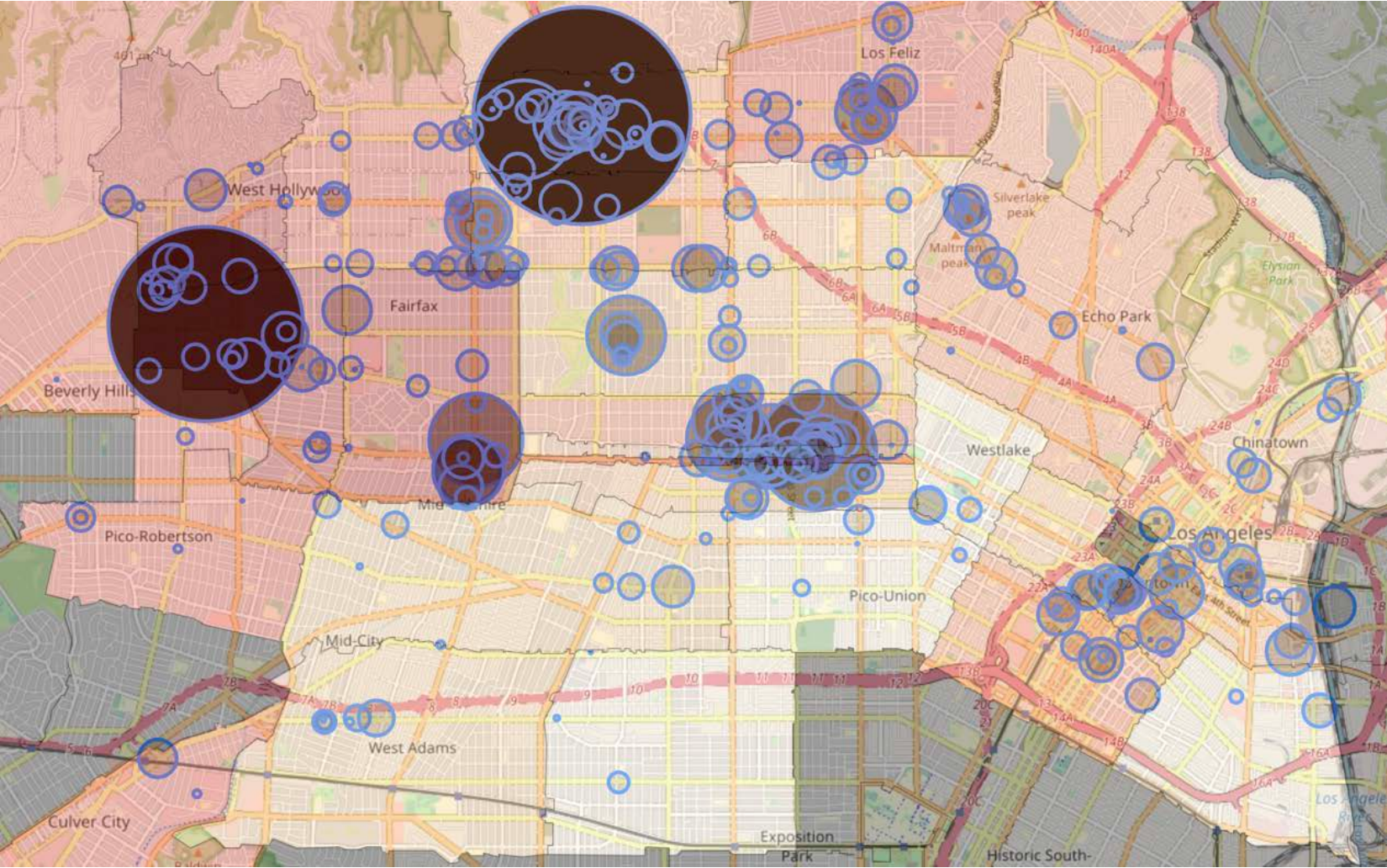} \label{fig:607}}
\subfloat[White Population]{\includegraphics[width=0.5\columnwidth]{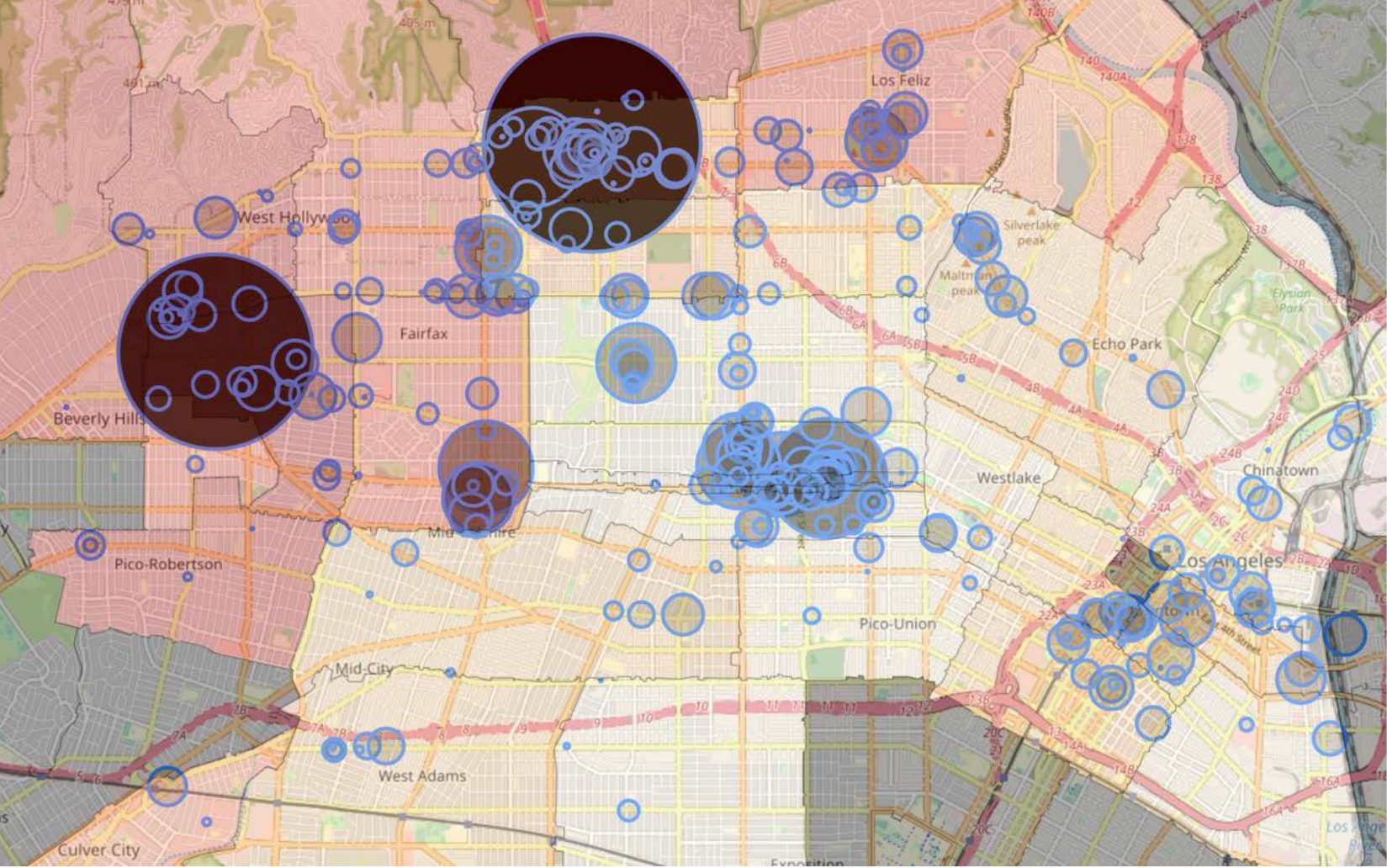} \label{fig:6055}} 
\caption{Heatmap for Los Angeles}
\label{LA-maps}
\end{figure}

\textbf{Los Angeles: } Figure~\ref{fig:LA} shows the exposure of restaurants in the city of Los Angeles. Using the DBSCAN algorithm, we were able to identify 172 clusters. 
Using our threshold, we obtained 8 hotspots. Table~\ref{log-yelp-LA-linearr} shows the results of our linear regression model. 
\begin{table}[htbp!] 
% \begin{minipage}{0.49\columnwidth}
\centering 
\caption{Results of the regression analysis of average \\exposure with hotspot for Los Angeles} 
  \label{log-yelp-LA-linearr} 
   \resizebox{0.6\columnwidth}{!}{%
\begin{tabular}{@{\extracolsep{5pt}}ll}
\hline 
& \multicolumn{1}{c}{\textit{Dependent variable:}} \\ 
\hline
\\[-1.8ex] & \multicolumn{1}{c}{\textit{average exposure}} \\ 
\\[-1.8ex] & \textit{Linear} \\ 
\hline \\[-1.8ex] 
Hotspot& 0.027 (0.009)$^{**}$ \\ 
  \hline \\[-1.8ex] 
Observations &344  \\ 
% Multiple R-squared & 0.01962 \\
Adjusted R-squared & 0.01676 \\
% F-statistic & 6.845 \\
\hline 
\textit{Note:}  & \multicolumn{1}{r}{$^{*}$p$<$0.05; $^{**}$p$<$0.01; $^{***}$p$<$0.001} \\ 
\end{tabular}
}
\end{table}
\begin{table}
\centering
% \begin{minipage}{0.35\columnwidth}
\caption{Results of the regression analysis of hotspot \\ with other sensitive attributes for Los Angeles} 
  \label{log-yelp-LA-log} 
   \resizebox{0.6\columnwidth}{!}{%
\begin{tabular}{@{\extracolsep{5pt}}ll}
\hline 
& \multicolumn{1}{c}{\textit{Dependent variable:}} \\ 
\hline
\\[-1.8ex] & \multicolumn{1}{c}{\textit{Hotspot}} \\ 
\\[-1.8ex] & \textit{Logistic} \\ 
\hline \\[-1.8ex] 
WN& - 0.387 (0.648) \\ 
BN& 0.088 (0.924) \\
AIN& 1.181 (0.186) \\
AN& 0.505 (0.408) \\
HED&16.627 (0.000)$^{***}$ \\
HUne& -2.087 (0.007) \\
HWe& 0.680 (0.541) \\
  \hline \\[-1.8ex] 
\textit{Note:}  & \multicolumn{1}{r}{$^{*}$p$<$0.05; $^{**}$p$<$0.01; $^{***}$p$<$0.001} \\ 
\end{tabular}
}
% \end{minipage}
\end{table} 
We can see similar results, that were obtained in the previous cities, where average exposure is positively correlated to hotspots of the city. Table~\ref{log-yelp-LA-log} shows the results of our logistic regression model. We see that restaurants in zip codes that are in a hotspot are in highly educated neighborhoods. Interestingly none of the demographic features were significant with hotspot. Figure~\ref{LA-maps} shows the heatmap for Los Angeles. We can see visually that major hotspots are in highly educated (shown in Figure~\ref{fig:607}) and restaurants with higher exposure are more in less white neighborhoods, hence confirming the validity of our regression analysis (shown in Figure~\ref{fig:6055}). It should also be noted that the areas which are highly educated are Beverly Crest and Westwood where they have about 67.6\% and 66.5\% adults who have earned a four-year degree or higher respectively.
Therefore, we can draw a conclusion about Yelp in the city of Los Angeles, businesses that are in the hotspots, have higher average exposure, supporting our \textit{H2}, and businesses that are in the zip codes that are in a hotspot, are in highly educated neighborhoods, hence we can reject \textit{H3}. 

\textbf{New Orleans: } Figure~\ref{fig:nola} shows the exposure of restaurants in the city of New Orleans. Using the DBSCAN algorithm, we were able to identify 87 clusters. 
Using our threshold, we obtained 3 hotspots. Table~\ref{log-yelp-NOLA-linearr} shows the results of our linear regression model. 
\begin{table}[htbp!] 
% \begin{minipage}{0.49\columnwidth}
\centering 
\caption{Results of the regression analysis of average \\exposure with hotspot for New Orleans} 
  \label{log-yelp-NOLA-linearr} 
   \resizebox{0.6\columnwidth}{!}{%
\begin{tabular}{@{\extracolsep{5pt}}ll}
\hline 
& \multicolumn{1}{c}{\textit{Dependent variable:}} \\ 
\hline
\\[-1.8ex] & \multicolumn{1}{c}{\textit{average exposure}} \\ 
\\[-1.8ex] & \textit{Linear} \\ 
\hline \\[-1.8ex] 
Hotspot& 0.039 (0.000)$^{***}$ \\ 
  \hline \\[-1.8ex] 
Observations &281  \\ 
% Multiple R-squared & 0.0482 \\
Adjusted R-squared & 0.04479\\
% F-statistic & 14.13 \\
\hline 
\textit{Note:}  & \multicolumn{1}{r}{$^{*}$p$<$0.05; $^{**}$p$<$0.01; $^{***}$p$<$0.001} \\ 
\end{tabular}
}
\end{table}
\begin{table}
\centering
% {0.49\columnwidth}
\caption{Results of the regression analysis of hotspot \\ with other sensitive attributes for New Orleans} 
  \label{log-yelp-NOLA-log} 
   \resizebox{0.6\columnwidth}{!}{%
\begin{tabular}{@{\extracolsep{5pt}}ll}
\hline 
& \multicolumn{1}{c}{\textit{Dependent variable:}} \\ 
\hline
\\[-1.8ex] & \multicolumn{1}{c}{\textit{Hotspot}} \\ 
\\[-1.8ex] & \textit{Logistic} \\ 
\hline \\[-1.8ex] 
WN& 7.261 (0.000)$^{***}$ \\ 
BN& 8.379 (0.000)$^{***}$ \\
AIN& -4.217 (0.000)$^{***}$ \\
AN& 0.754 (0.329) \\
HED& 4.067 (0.000)$^{***}$ \\
HUne& 3.296 (0.000)$^{***}$ \\
HWe& 1.217 (0.140) \\
  \hline \\[-1.8ex] 
\textit{Note:}  & \multicolumn{1}{r}{$^{*}$p$<$0.05; $^{**}$p$<$0.01; $^{***}$p$<$0.001} \\ 
\end{tabular}
}
% \end{minipage}
\end{table} 
We can see similar results, that were obtained in the previous cities, where average exposure is positively correlated to the hotspots of the city. Table~\ref{log-yelp-NOLA-log} shows the results of our logistic regression model. We see that restaurants in zip codes that are in a hotspot are in highly white, black neighborhoods, they are also in highly educated and highly unemployed areas. This is interesting, however, New Orleans had the highest unemployment rate among large metro areas according to U.S. Bureau of Labor Statistics~\cite{nola-power}.
\begin{figure}[htbp!]
\centering
\subfloat[Education] {\includegraphics[width=0.5\columnwidth]{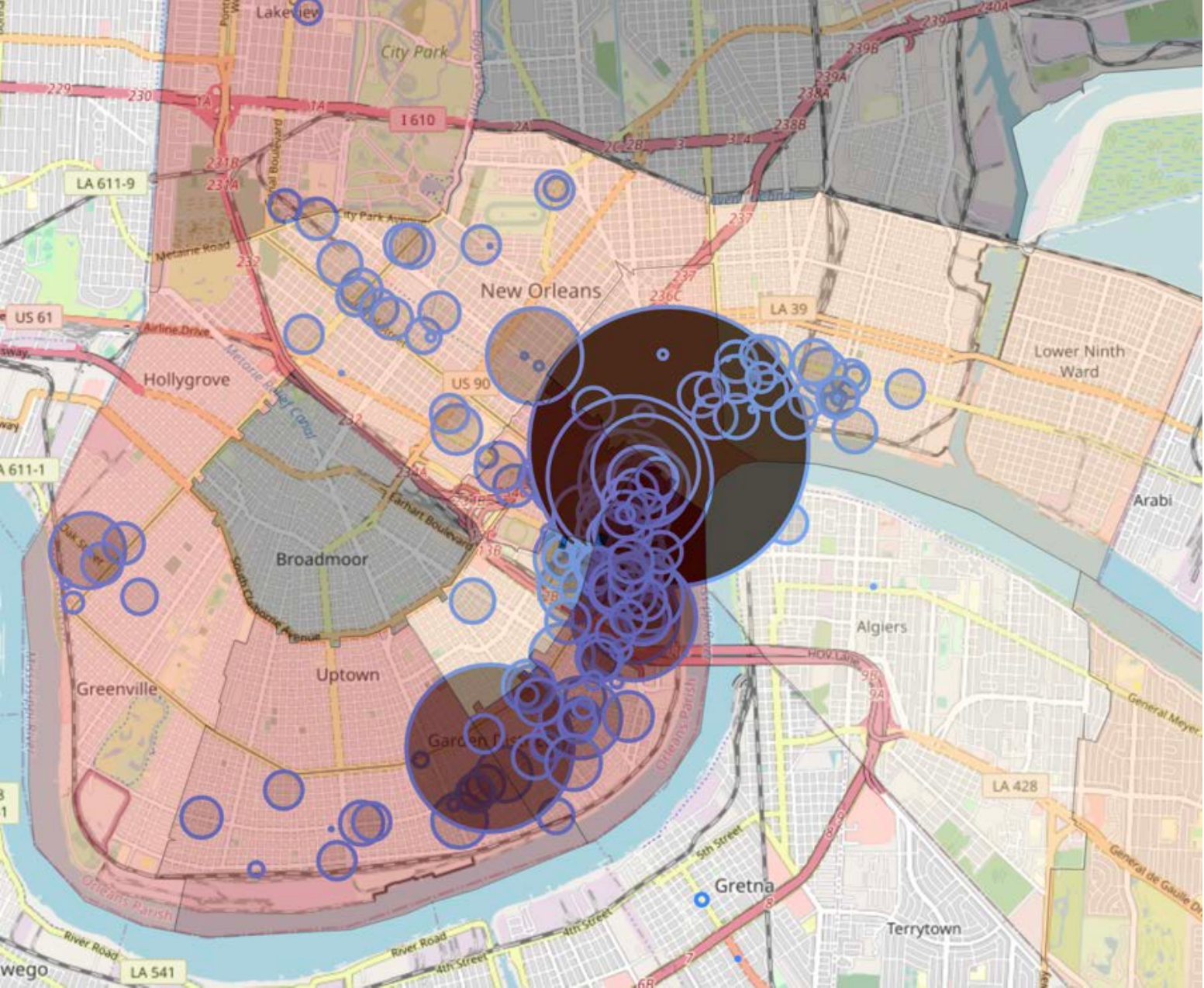} \label{fig:60778}}
\subfloat[Unemployment]{\includegraphics[width=0.5\columnwidth]{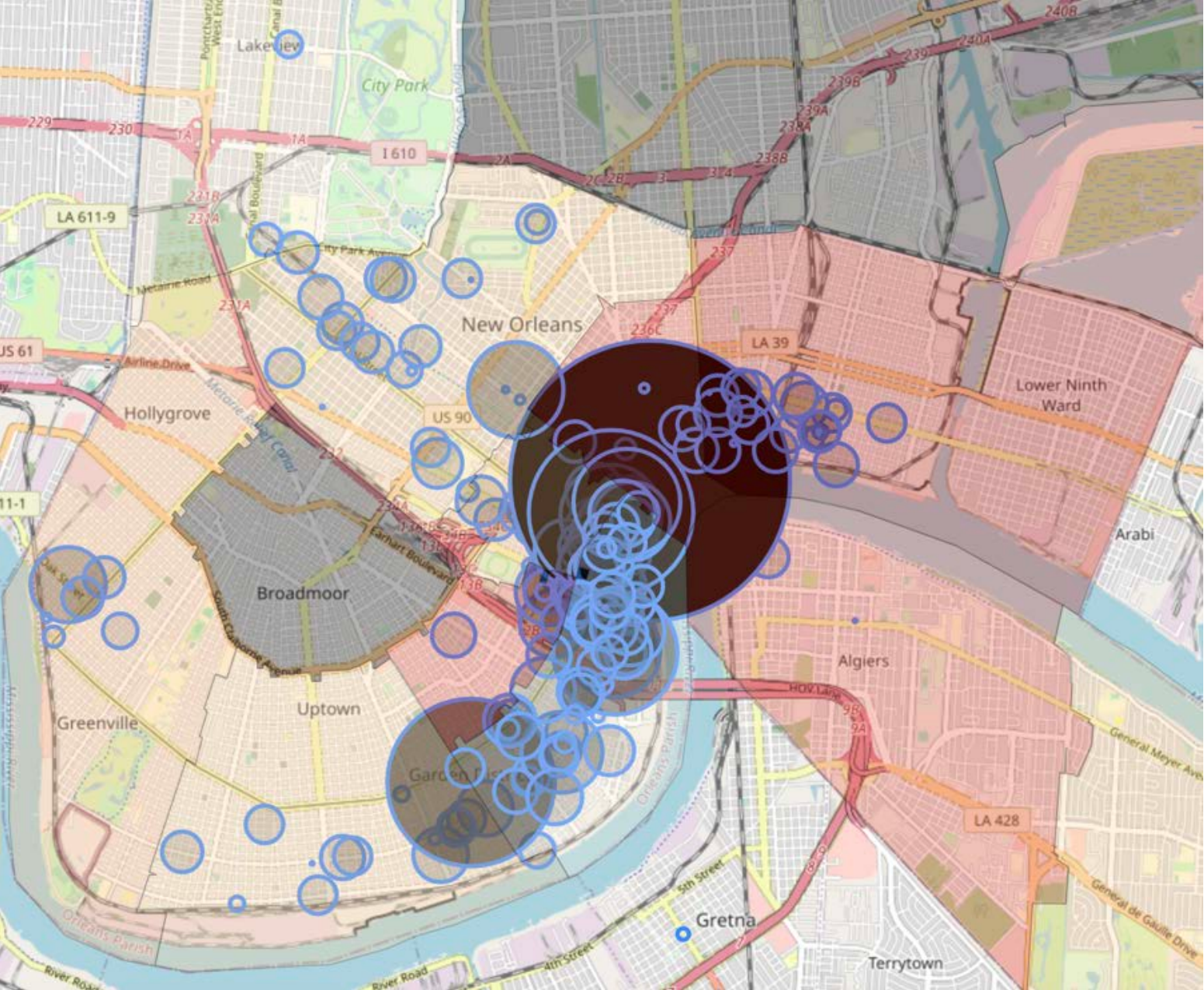} \label{fig:60w55}} 
\caption{Heatmap for New Orleans}
\label{NOLA-maps}
\end{figure}
Figure~\ref{NOLA-maps} shows the heatmap for New Orleans. We can see visually that major hotspots are in highly educated (shown in Figure~\ref{fig:60778}) and highly unemployed neighborhoods, (shown in Figure~\ref{fig:60w55}).
Therefore, we can draw a conclusion about Yelp in the city of New Orleans, businesses that are in hotspots, have higher average exposure, hence finding support for \textit{H2}, and businesses in zip codes where there is a hotspot, are in highly educated and highly unemployed neighborhoods and diverse neighborhoods, hence we find partial support for \textit{H3}.

\textbf{New York City: } Figure~\ref{fig:nyc} shows the exposure of restaurants in the city of New York City. 
Using the DBSCAN algorithm, we were able to identify 131 clusters. 
Using our threshold, we obtained 9 hotspots in the city of New York. Table~\ref{log-yelp-NYC-linearr} shows the results of our linear regression model. 

\begin{table}[htbp!]
\centering 
\caption{Results of the regression analysis of average \\ exposure with hotspot for New York City} 
  \label{log-yelp-NYC-linearr} 
   \resizebox{0.6\columnwidth}{!}{%
\begin{tabular}{@{\extracolsep{5pt}}ll}
\hline 
& \multicolumn{1}{c}{\textit{Dependent variable:}} \\ 
\hline
\\[-1.8ex] & \multicolumn{1}{c}{\textit{average exposure}} \\ 
\\[-1.8ex] & \textit{Linear} \\ 
\hline \\[-1.8ex] 
Hotspot& 0.031 (0.001)$^{**}$ \\ 
  \hline \\[-1.8ex] 
Observations &348  \\ 
% Multiple R-squared & 0.02823 \\
Adjusted R-squared & 0.02542\\
% F-statistic & 10.05 \\
\hline 
\textit{Note:}  & \multicolumn{1}{r}{$^{*}$p$<$0.05; $^{**}$p$<$0.01; $^{***}$p$<$0.001} \\ 
\end{tabular}
}
\end{table}
% hfill
\begin{table}[htbp!]
% {0.49\columnwidth}
\centering
\caption{Results of the regression analysis of hotspot \\with other sensitive attributes for New York} 
  \label{log-yelp-NYC-log} 
   \resizebox{0.6\columnwidth}{!}{%
\begin{tabular}{@{\extracolsep{5pt}}ll}
\hline 
& \multicolumn{1}{c}{\textit{Dependent variable:}} \\ 
\hline
\\[-1.8ex] & \multicolumn{1}{c}{\textit{Hotspot}} \\ 
\\[-1.8ex] & \textit{Logistic} \\ 
\hline \\[-1.8ex] 
WN& -2.077 (0.018)$^{*}$ \\ 
BN& -2.046 (0.027)$^{*}$ \\
AIN& 0.597 (0.359)\\
AN& 1.218 (0.031)$^{*}$ \\
HED&N/A \\
HUne& 0.186 (0.749)\\
HWe& 2.197 (0.018)$^{*}$ \\
  \hline \\[-1.8ex] 
\textit{Note:}  & \multicolumn{1}{r}{$^{*}$p$<$0.05; $^{**}$p$<$0.01; $^{***}$p$<$0.001} \\ 
\end{tabular}
}
% \end{minipage}
\end{table} 

We can see similar results, that were obtained in the previous cities, where average exposure is positively correlated to hotspots of the city. Table~\ref{log-yelp-NYC-log} shows the results of our logistic regression model. We see that restaurants in zip codes that are in a hotspot are highly Asian and highly wealthy neighborhoods. We also observed that restaurants in zip codes that are in a hotspot have lower numbers of white and black populations. Interestingly, we didn't find any results for education. 
\begin{figure}[htbp!]
\centering
\subfloat[Education] {\includegraphics[width=0.45\columnwidth]{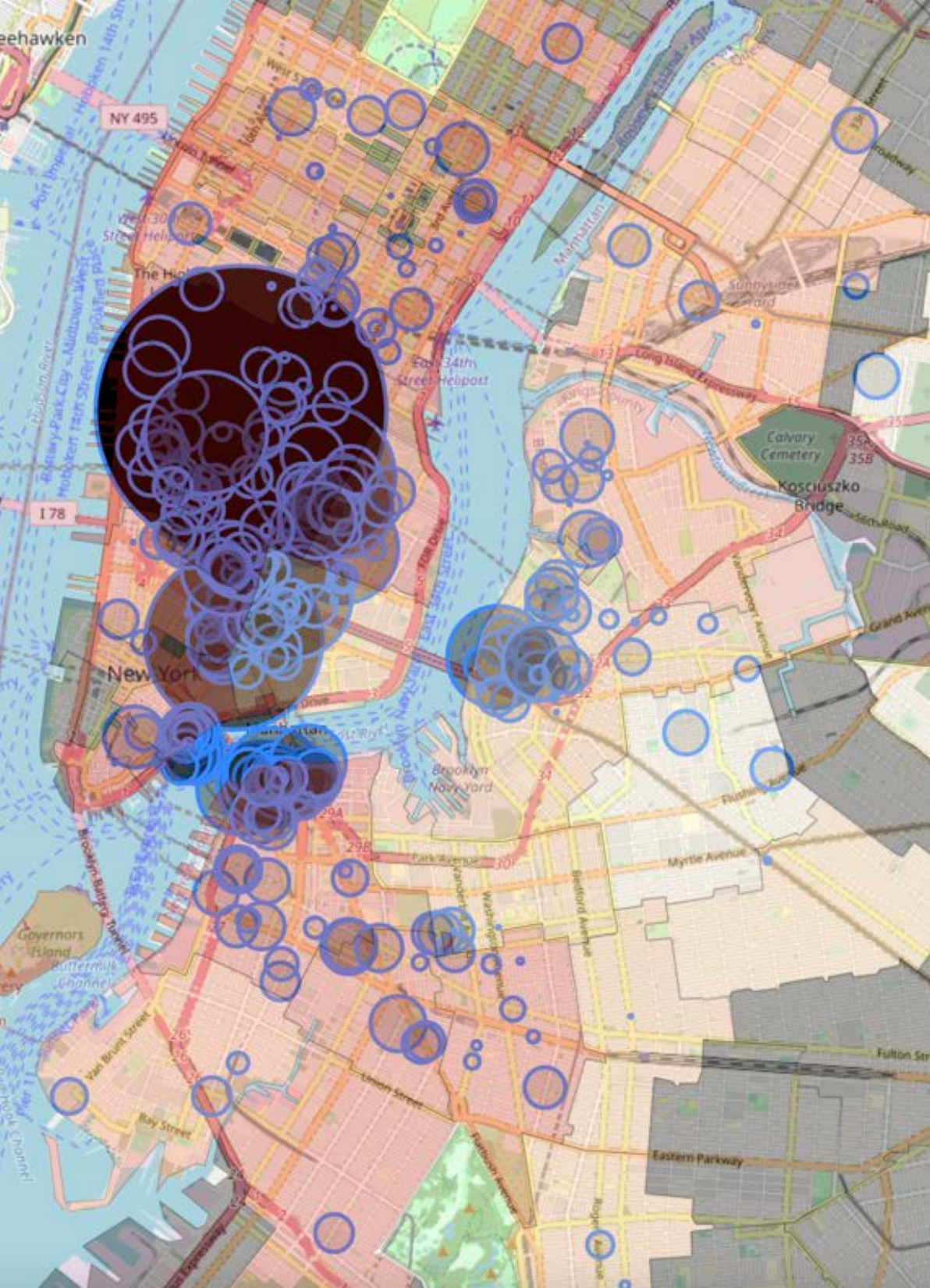} \label{fig:60779}}
\subfloat[Wealth]{\includegraphics[width=0.45\columnwidth]{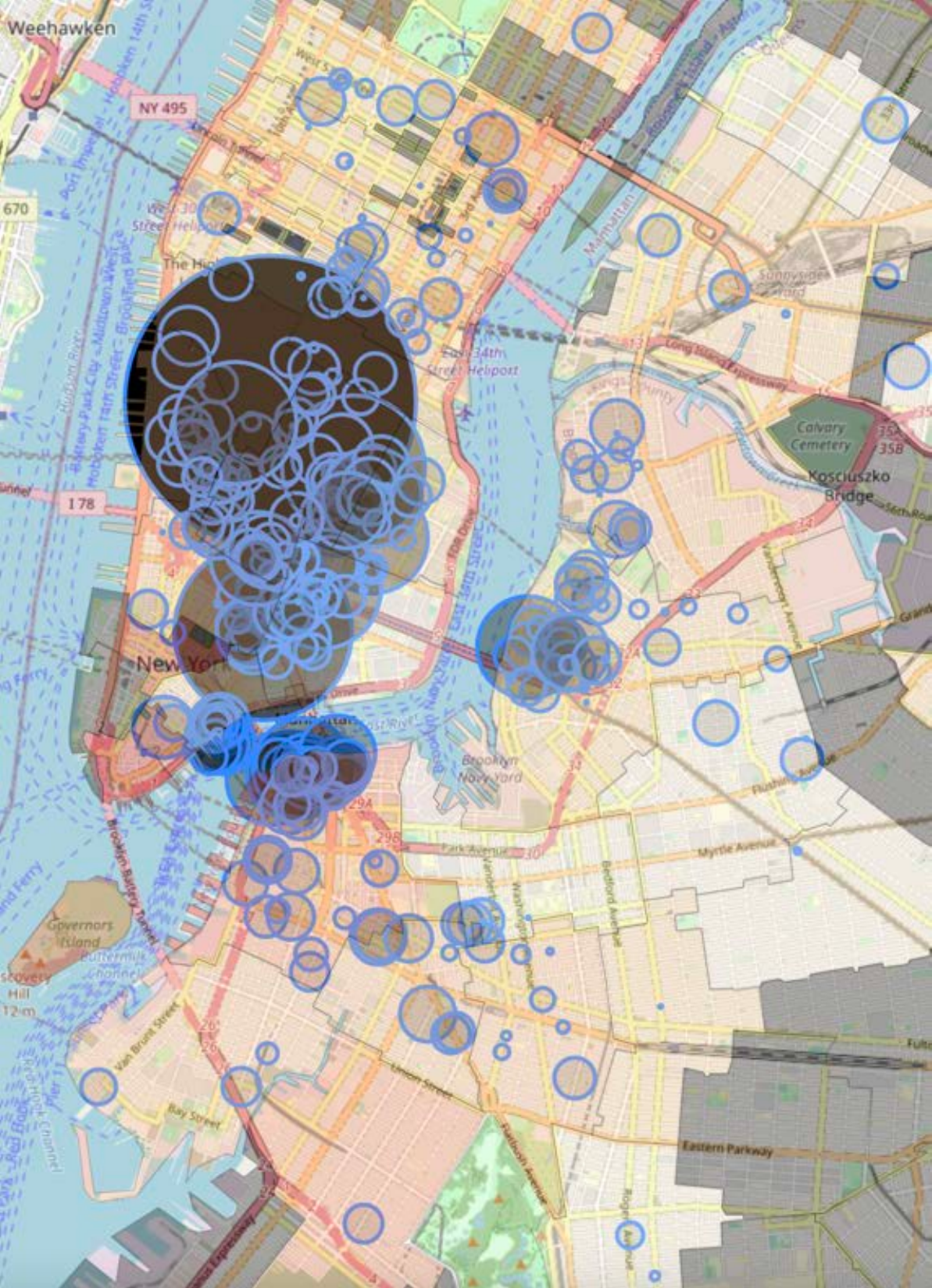} \label{fig:60w155}} 
\caption{Heatmap for New York City}
\label{NYC-maps}
\end{figure}
Figure~\ref{NYC-maps} shows the heatmap for New York City. We can visually see that major hotspots are all scattered in highly educated neighborhoods(shown in Figure.~\ref{fig:60779}), hence confirming why we did not obtain any result from our regression model and highly wealthy neighborhoods, (shown in Figure~\ref{fig:60w155}).
Therefore, we can draw a conclusion about Yelp in the city of New York, businesses that are in the hotspots, have higher average exposure, hence finding support for \textit{H2}, and businesses that are in zip codes that are in a hotspot, are in highly wealthy and Asian neighborhoods and they are also in the neighborhoods that have lower white and black populations, hence we partially find support for \textit{H3}.

\textbf{San Francisco: } Figure~\ref{fig:sanfan} shows the exposure of restaurants in the city of San Francisco. Using the DBSCAN algorithm, we were able to identify 120 clusters. 
Using our threshold, we obtained 13 hotspots in the city of San Francisco. Table~\ref{log-yelp-SAFO-linearr} shows the results of our linear regression model. 
\begin{table}[htbp!]
% \begin{minipage}{0.49\columnwidth}
\centering 
\caption{Results of the regression analysis of average \\ exposure with hotspot for San Francisco} 
  \label{log-yelp-SAFO-linearr} 
   \resizebox{0.6\columnwidth}{!}{%
\begin{tabular}{@{\extracolsep{5pt}}ll}
\hline 
& \multicolumn{1}{c}{\textit{Dependent variable:}} \\ 
\hline
\\[-1.8ex] & \multicolumn{1}{c}{\textit{average exposure}} \\ 
\\[-1.8ex] & \textit{Linear} \\ 
\hline \\[-1.8ex] 
Hotspot& 0.021 (0.020)$^{*}$ \\ 
  \hline \\[-1.8ex] 
Observations &353 \\ 
% Multiple R-squared & 0.01519 \\
Adjusted R-squared & 0.01239\\
% F-statistic & 5.415 \\
\hline 
\textit{Note:}  & \multicolumn{1}{r}{$^{*}$p$<$0.05; $^{**}$p$<$0.01; $^{***}$p$<$0.001} \\ 
\end{tabular}
}
\end{table}
% \hfill
\begin{table}[htbp!]
\centering
% {0.49\columnwidth}
\caption{Results of the regression analysis of hotspot \\with other sensitive attributes for San Francisco} 
  \label{log-yelp-SAFO-log} 
   \resizebox{0.6\columnwidth}{!}{%
\begin{tabular}{@{\extracolsep{5pt}}ll}
\hline 
& \multicolumn{1}{c}{\textit{Dependent variable:}} \\ 
\hline
\\[-1.8ex] & \multicolumn{1}{c}{\textit{Hotspot}} \\ 
\\[-1.8ex] & \textit{Logistic} \\ 
\hline \\[-1.8ex] 
WN& -1.457 (0.144) \\ 
BN& 0.027 (0.963) \\
AIN& -0.786 (0.303) \\
AN& -1.812 (0.225) \\
HED&13.214 (0.000)$^{***}$ \\
HUne& 0.826 (0.368) \\
HWe& 1.491 (0.049)$^{*}$ \\
  \hline \\[-1.8ex] 
\textit{Note:}  & \multicolumn{1}{r}{$^{*}$p$<$0.05; $^{**}$p$<$0.01; $^{***}$p$<$0.001} \\ 
\end{tabular}
}
% \end{table}
\end{table} 
We can see similar results, that were obtained in the previous cities, where average exposure is positively correlated to the hotspots of the city. Table~\ref{log-yelp-SAFO-log} shows the results of our logistic regression model. We see that restaurants in zip codes that are in a hotspot are in highly educated and are in highly wealth neighbourhoods. Interestingly none of the demographic groups have any correlation with hotspots. 

\begin{figure}[htbp!]
\centering
\subfloat[Education] {\includegraphics[width=0.5\columnwidth]{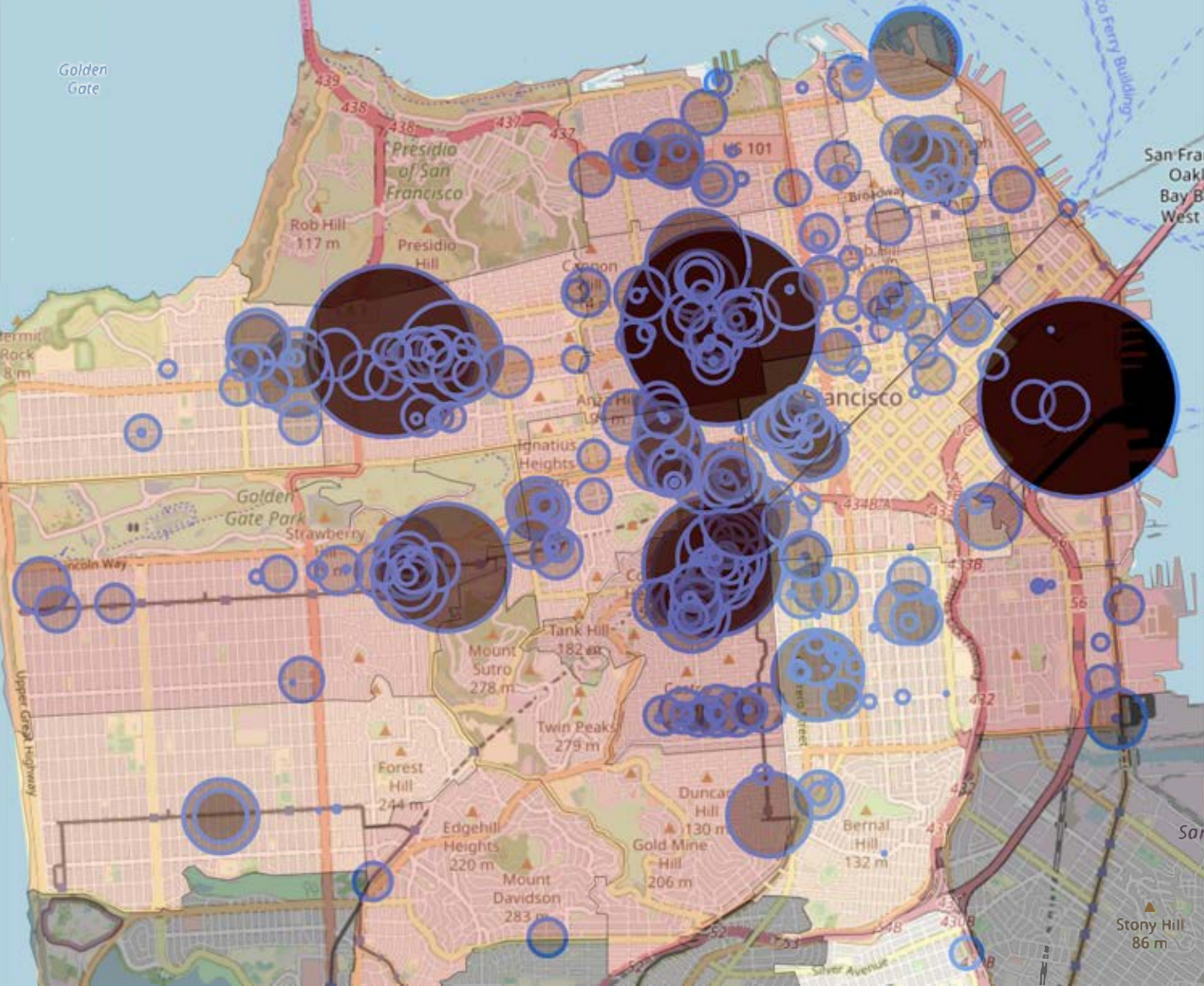} \label{fig:60780}}
\subfloat[Wealth]{\includegraphics[width=0.5\columnwidth]{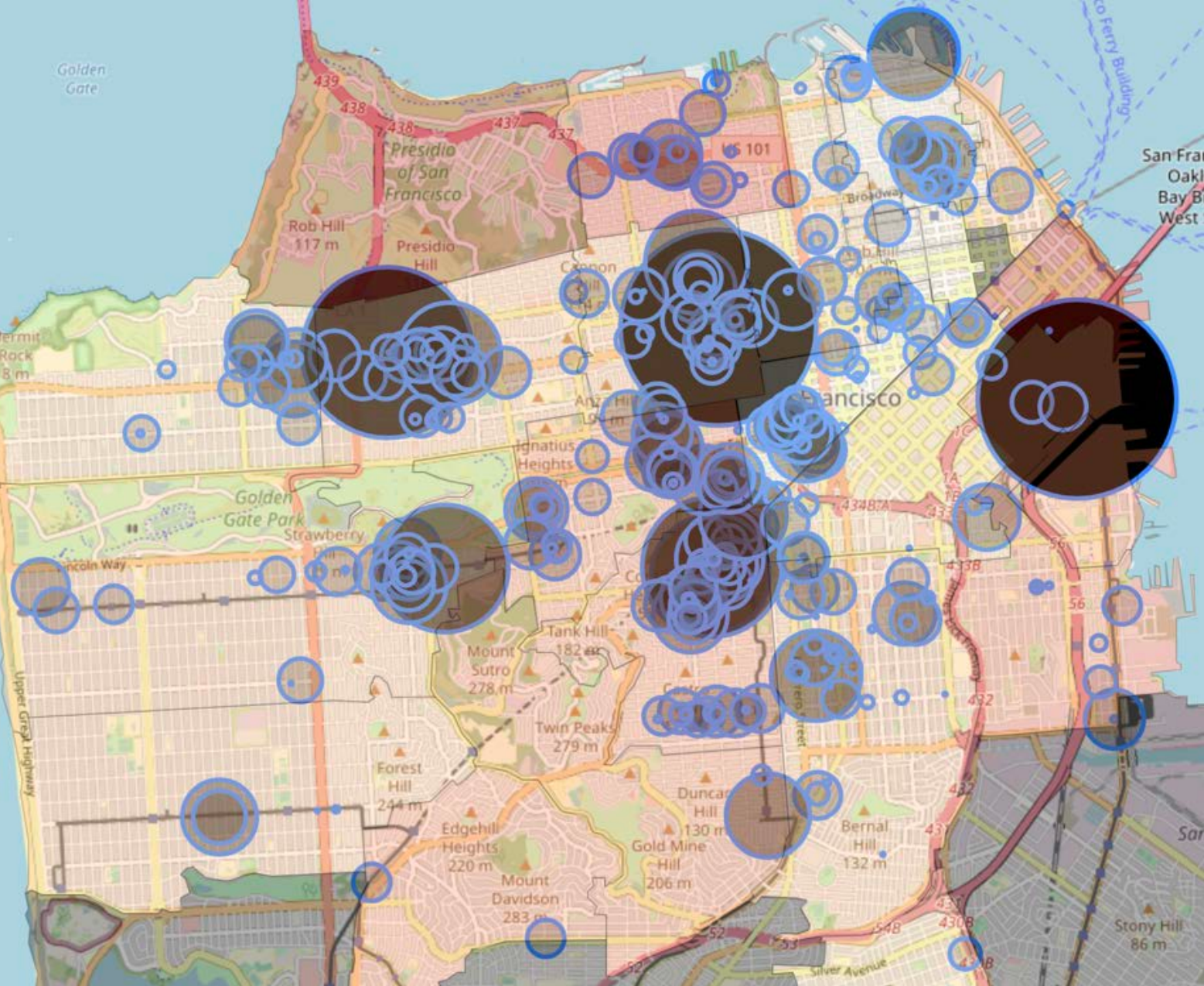} \label{fig:60w156}} 
\caption{Heatmap for San Francisco}
\label{SAFO-maps}
\end{figure}
Figure~\ref{SAFO-maps} shows the heatmap for San Francisco. We can visually see that major hotspots are all scattered in highly educated (shown in Figure~\ref{fig:60780}) and highly wealthy neighborhoods, (shown in Figure~\ref{fig:60w156}).
Therefore, we can draw a conclusion about Yelp in the city of San Francisco, businesses that are in the hotspots, have higher average exposure, and businesses that are in zip codes that are in a hotspot, are in highly wealthy and highly educated neighborhoods. While we do find support for \textit{H2}, we were only able to partially find support for \textit{H3}. 

\textbf{San Jose: } Figure~\ref{fig:sanjose} shows the exposure of restaurants in the city of San Jose. Using the DBSCAN algorithm, we were able to identify 151 clusters. 
Using our threshold, we obtained 10 hotspots in the city of San Jose. Table~\ref{log-yelp-SJC-linearr} shows the results of our linear regression model. 
\begin{table}[htbp!]
\centering 
\caption{Results of the regression analysis of average \\ exposure with hotspot for San Jose} 
  \label{log-yelp-SJC-linearr} 
   \resizebox{0.6\columnwidth}{!}{%
\begin{tabular}{@{\extracolsep{5pt}}ll}
\hline 
& \multicolumn{1}{c}{\textit{Dependent variable:}} \\ 
\hline
\\[-1.8ex] & \multicolumn{1}{c}{\textit{average exposure}} \\ 
\\[-1.8ex] & \textit{Linear} \\ 
\hline \\[-1.8ex] 
Hotspot& 0.027 (0.005)$^{**}$ \\ 
  \hline \\[-1.8ex] 
Observations &346  \\ 
% Multiple R-squared & 0.01519 \\
Adjusted R-squared & 0.01239\\
% F-statistic & 5.415 \\
\hline 
\textit{Note:}  & \multicolumn{1}{r}{$^{*}$p$<$0.05; $^{**}$p$<$0.01; $^{***}$p$<$0.001} \\ 
\end{tabular}
}
\end{table} 
We can see similar results, that were obtained in the previous cities, where average exposure is positively correlated to hotspots of the city, hence we find support for \textit{H2}. 

\textbf{Seattle: } Figure~\ref{fig:seat} shows the exposure of restaurants in the city of Seattle. Using the DBSCAN algorithm, we were able to identify 103 clusters. 
Using our threshold, we obtained 11 hotspots in the city of San Jose. Table~\ref{log-yelp-SWA-linearr} shows the results of our linear regression model. 
\begin{table}[htbp!]
% \begin{minipage}{0.49\columnwidth}
\centering 
\caption{Results of the regression analysis of average \\ exposure with hotspot for Seattle} 
  \label{log-yelp-SWA-linearr} 
   \resizebox{0.6\columnwidth}{!}{%
\begin{tabular}{@{\extracolsep{5pt}}ll}
\hline 
& \multicolumn{1}{c}{\textit{Dependent variable:}} \\ 
\hline
\\[-1.8ex] & \multicolumn{1}{c}{\textit{average exposure}} \\ 
\\[-1.8ex] & \textit{Linear} \\ 
\hline \\[-1.8ex] 
Hotspot& 0.023 (0.025)$^{*}$ \\ 
  \hline \\[-1.8ex] 
Observations &305  \\ 
% Multiple R-squared & 0.01637 \\
Adjusted R-squared & 0.01312\\
% F-statistic & 5.042 \\
\hline 
\textit{Note:}  & \multicolumn{1}{r}{$^{*}$p$<$0.05; $^{**}$p$<$0.01; $^{***}$p$<$0.001} \\ 
\end{tabular}
}
\end{table}
% \hfill

\begin{table}[htbp!]
\centering
% {0.49\columnwidth}
\caption{Results of the regression analysis of hotspot \\with other sensitive attributes for Seattle} 
  \label{log-yelp-SWA-log} 
   \resizebox{0.6\columnwidth}{!}{%
\begin{tabular}{@{\extracolsep{5pt}}ll}
\hline 
& \multicolumn{1}{c}{\textit{Dependent variable:}} \\ 
\hline
\\[-1.8ex] & \multicolumn{1}{c}{\textit{Hotspot}} \\ 
\\[-1.8ex] & \textit{Logistic} \\ 
\hline \\[-1.8ex] 
WN& -3.145 (0.092) \\ 
BN& -0.122 (0.933) \\
AIN& -2.629 (0.016)$^{*}$ \\
AN& -0.709 (0.659)\\
HED&-0.441 (0.795) \\
HUne& 0.059 (0.965) \\
HWe& -2.066 (0.000)$^{***}$ \\
  \hline \\[-1.8ex] 
\textit{Note:}  & \multicolumn{1}{r}{$^{*}$p$<$0.05; $^{**}$p$<$0.01; $^{***}$p$<$0.001} \\ 
\end{tabular}
}
% \end{minipage}
\end{table} 
We can see similar results, that were obtained in the previous cities, where average exposure is positively correlated to hotspots of the city. Table~\ref{log-yelp-SWA-log} shows the results of our logistic regression model. We find that restaurants in zip codes that are in a hotspot have a smaller American Indian population and are in less wealthy neighborhoods. 
\begin{figure}[htbp!]
\centering
\includegraphics[width=0.45\columnwidth]{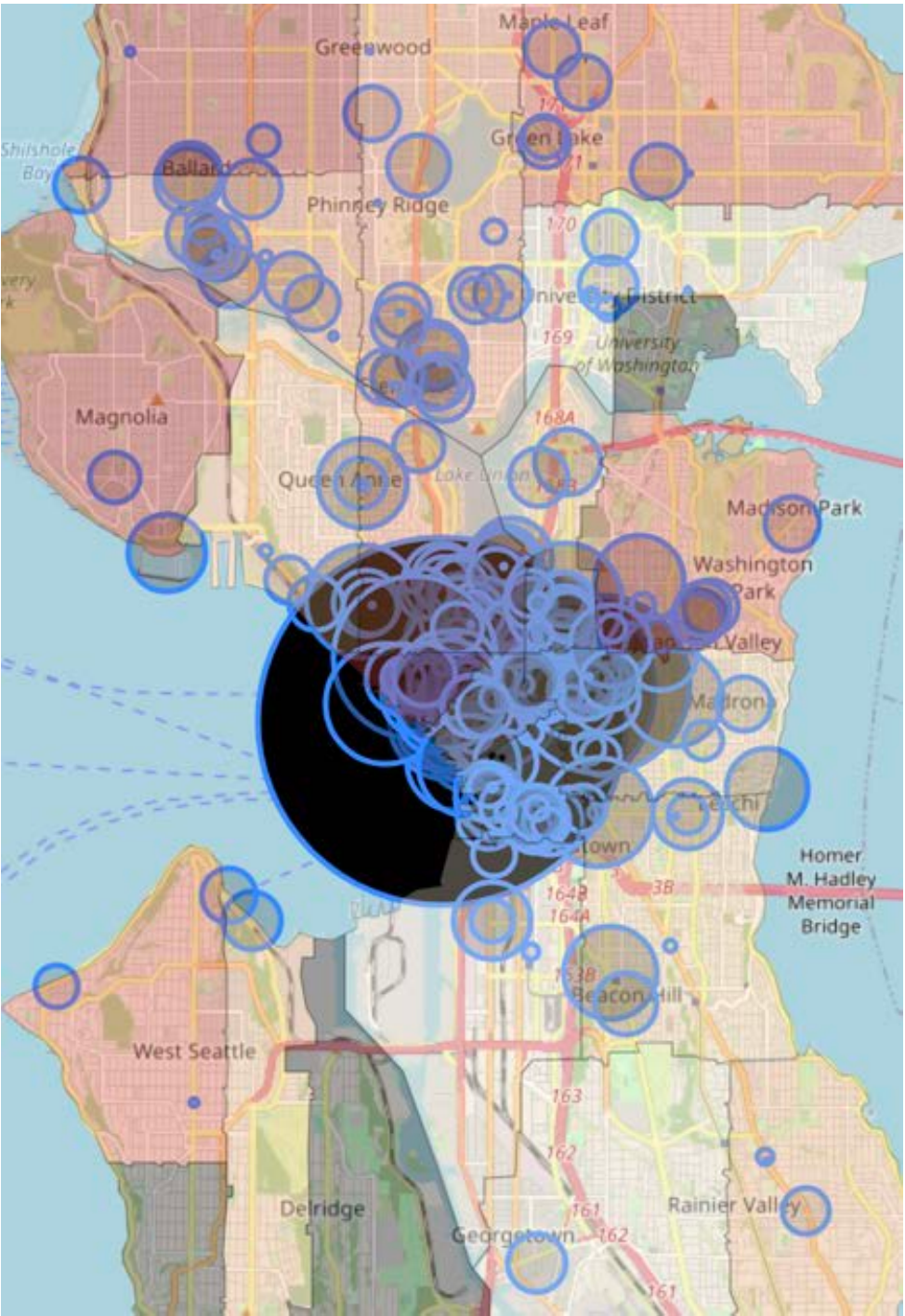} 
\label{fig:60w1544}
\caption{Wealth heatmap for Seattle}
\label{SWA-maps}
\end{figure}

Figure~\ref{SWA-maps} shows the heatmap of wealth for the city of Seattle. We can clearly see that the hotspots are in areas where the wealth is less, hence our analysis stands true.
Therefore, we can draw a conclusion about Yelp in the city of Seattle, businesses that are in the hotspots, have higher average exposure, and businesses that are in zip codes that are in a hotspot, are in less wealthy neighborhoods and have less number of American Indian population. While we find support for \textit{H2}, we reject our \textit{H3}.

\end{document}